\documentclass[twocolumn,preprintnumbers,amsmath,amssymb]{revtex4}

\usepackage{color}
\usepackage{mathptmx}
\usepackage{amsmath} 
\usepackage{amssymb} 
\usepackage{graphicx}
\usepackage{dcolumn}
\usepackage{bm}
\newcommand{\mean}[1]{\langle#1\rangle}
\newcommand{\ket}[1]{\mid#1\rangle}
\newcommand{\bra}[1]{\langle#1\mid} 
\newcommand{\ud}[1]{#1^\dagger}
\newcommand{\Tr}[1]{\mathrm{Tr}(#1)}

\begin{document}


\preprint{APS/123-QED}

\title{Strong and weak coupling of two coupled qubits}

\author{Elena del Valle} \email{elena.delvalle.reboul@gmail.com}
\affiliation{School of Physics and Astronomy, University of
  Southampton, SO17 1BJ, Southampton, United Kingdom}

\date{\today}

\begin{abstract}
  I investigate the dynamics and power spectrum of two coupled qubits
  (two-level systems) under incoherent continuous pump and
  dissipation.  New regimes of strong coupling are identified, that
  are due to additional paths of coherence flow in the system.
  Dressed states are reconstructed even in the regime of strong
  decoherence. The results are analytical and offer an exact
  description of strong-coupling in presence of pumping and decay in a
  nontrivial (nonlinear) system.
\end{abstract}

\maketitle
\section{Introduction}

Quantum information processing~\cite{nielsen_book00a} presupposes a
coherent coupling of the fundamental bricks of quantum information,
the qubits. In real systems, however, decoherence, dissipation and
incoherent coupling to the environment is
unavoidable~\cite{gardiner_book00a}. In the field of cavity quantum
electrodynamics~\cite{haroche_book06a}, the notion of coherent
coupling is known as \emph{strong coupling}, in this case, between
light and matter (e.g., the excited state of an atomic
transition~\cite{allen_book87a} or an electron-hole pair in a
semiconductor~\cite{kavokin_book07a}). This leads to a quantum
superposition of the bare states, resulting in so-called \emph{dressed
  states}~\cite{cohentannoudji_book01a}. This regime is reached in
systems of very high quality and under tight experimental control, so
that intrinsic sources of decoherence are minimized as much as
possible and coherent dynamics takes over. The simplest description of
strong coupling neglects dissipation altogether and thus reduces to
that of mere coupling with strength~$g$, introducing the notion of
\emph{Rabi splitting}~\cite{sanchezmondragon83a}.  Next step in the
description includes the decay $\gamma_i$ of the bare states, $i=1$,
2~\cite{carmichael89a}. This gives rise to a widely known criterion
for strong-coupling: $4g>|\gamma_1-\gamma_2|$. This is the case of
vacuum Rabi splitting where at most one excitation is involved. At
this level, which is the most natural and fundamental since it
describes one particle, there is no difference from the underlying
theoretical model. Differences appear at the next step of description
when the excitation scheme is taken into account. A typical
description of excitation is to consider an initial condition. The
coupling is then studied as the spontaneous emission from this initial
excited state. Unless the initial condition is restricted to one
excitation (as previously), the underlying theoretical model becomes
determinant. Another important description of the excitation process
is that of a continuous pumping, for instance a coherent excitation
that drives the system~\cite{mollow69a,cohentannoudji77a}, or an
incoherent pump that feeds excitations at a given rate but without any
coherent input~\cite{cirac91a,tian92a}. The latter is more directly
related to the intrinsic dynamics of the system, and is the one that
will be considered in this text. With non-negligible pumping, the
underlying theoretical description cannot be ignored, and taking it
into account leads to strong deviations from the paradigm of
strong-coupling as established by the spontaneous emission of one
excitation (in any model)~\cite{delvalle_book09a}.

\begin{figure}[t]
\centering
\includegraphics[width=\linewidth]{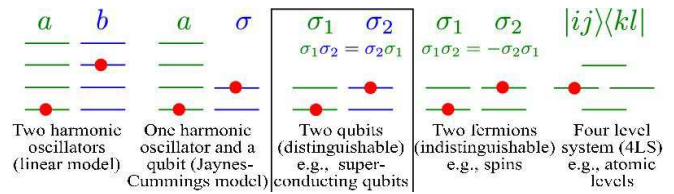}
\caption{(Color online) Schema of the systems of interest in this
  text. The main object of study, two coupled qubits, is sketched in
  (c). It will be compared throughout with other coupled systems, (a),
  (b), (d) and (e).}
\label{fig:SatJan9162053GMT2010}
\end{figure}

An immediate extension of light-matter coupling in the linear regime
is the \emph{linear model}, namely, that of two harmonic oscillators
with no restriction in the number of particles
Fig.~\ref{fig:SatJan9162053GMT2010}(a). Its comprehensive description
under incoherent pumping was given in Ref.~\cite{laussy09a}. This
describes for instance exciton-polaritons in planar semiconductor
microcavities~\cite{kavokin_book07a} (where both excitons and photons
are bosons). It was shown in this work how pumping calls for extended
definitions of strong-coupling, essentially requiring absorption of
the pumping rates in the decay rates.  A more important theoretical
model, known as the Jaynes-Cummings model~\cite{jaynes63a}, describes
the coupling of a two-level system (such as an atom or a
zero-dimensional exciton in a small quantum dot) with a boson mode
(typically, cavity photons),
Fig.~\ref{fig:SatJan9162053GMT2010}(b). It is more important because
more closely related to a genuinely quantum regime, the linear model
being essentially a classical description cast in quantum-mechanical
terms. Its description under incoherent pumping was given in
Ref.~\cite{delvalle09a}, but encountered various difficulties to offer
a complete picture. In particular, dressed modes exhibit complex
patterns and a definition of strong-coupling in this system is much
more difficult to achieve, since splitting of the dressed states
depends on the excitations. In particular, strong-coupling can be
enforced by pumping, leading to situations of mixed weak and strong
coupling, where some of the states are bare while some others are
dressed.

In this text, I will address the case of two two-level systems
(Fig.~\ref{fig:SatJan9162053GMT2010}(cde)), which compromises between
simplicity of the linear model and richness of the Jaynes-Cummings
model. In particular, thanks to the reduced size of the Hilbert space,
I will be able to solve the problem fully analytically, as in the
linear model, a convenience not afforded by the Jaynes-Cummings model
(when including incoherent pumping). This will allow me to provide a
complete picture of weak and strong coupling in a nontrivial system,
and therefore shed light on more complicated systems.

My description will address more particularly independent qubits
(Fig.~\ref{fig:SatJan9162053GMT2010}(c)), for instance superconducting
(Josephson) qubits~\cite{pashkin09a}, in the sense that their
commutation rules will not be those of two fermions, that anticommute
(Fig.~\ref{fig:SatJan9162053GMT2010}(d)). I will also address the
latter case for comparison and completeness, and obtain the elegant
result that the expressions describing two coupled fermions are
essentially identical to those describing two coupled bosons, although
these two systems are very different in character and behavior (for
instance bosons accumulate arbitrary number of particles whereas
fermions saturate at at most one, a distinction recovered in the
formalism by merely substituting effective parameters). Also because
two two-level systems can be mapped to one four-level system
(Fig.~\ref{fig:SatJan9162053GMT2010}(e)), I will address this case in
detail, finding another fundamental case of interest.

The main results, however, and the deepest connections to be made with
other models (such as the Jaynes-Cummings) will be obtained from the
case of two qubits. Beyond its interest for the previous reasons, the
study of the coupling between two qubits is interesting in its own
right~\cite{pashkin03a,majer05a,hime06a,niskanen07a}: it comes as the
fundamental support of entangled states~\cite{wootters98a,berkley03a},
to implement quantum gates~\cite{yamamoto03a,li03a} and, in this
quantum information processing context, the natural model to
investigate decoherence~\cite{grigorenko05a, galperin05a}. In this
text, the two coupled qubits will allow us to investigate strong
coupling under incoherent pumping.

The rest of this paper is organized as follows. In
section~\ref{sec:MonFeb16112240CET2009}, I introduce the model and its
parameters and discuss the level structure. In
section~\ref{sec:SunFeb1051813CET2009}, I obtain the single-time
dynamics, that will be shown to be the same independently of the
underlying model. In section \ref{sec:WedJul29174809GMT2009}, I obtain
the power spectra, that, in contrast with the previous section, depend
strikingly on the underlying model.  Power spectra (photoluminescence
spectra in a quantum optical context) are important because this is
where the dressed states manifest. I will analyze in detail the case
of two qubits (Sec.~\ref{sec:SunJan10204219GMT2010}) and contrast it
with that of two fermions (Sec.~\ref{sec:FriNov20155955WET2009}) and
of a four-level system (Sec.~\ref{sec:MoNov23201725WET2009}).  In
section~\ref{sec:MonFeb16112450CET2009}, I describe the strong and
weak coupling regimes, first in the absence
(Sec.~\ref{sec:SunJan10204807GMT2010}), and then including
(\ref{sec:FriFeb6020814CET2009}) the incoherent continuous pump. I
define new regimes of strong coupling, proper to the two qubits
system. In section~\ref{sec:SunFeb1004214CET2009}, I illustrate the
results of previous sections with examples of some interesting
configurations: cases where pumping effect is optimal
(Sec.~\ref{sec:WedOct14101637GMT2009}) and detrimental
(Sec.~\ref{sec:WedOct14131208GMT2009}) for the coherent coupling.  In
section~\ref{sec:WedJul29173348GMT2009}, based on the previous
results, I reconstruct the dressed states, uncovering an unexpected
manifestation of decoherence---the emergence of additional dressed
states---to be found only in the coexistence of pumping \emph{and}
decay along with the coherent coupling. Finally, in
section~\ref{sec:SatMay19212524CEST2007}, I give a summary of the main
results as my conclusions. To avoid distraction in the main text, most
technical details appear in appendices
(\ref{app:MonJul6000455GMT2009}, \ref{app:WedJul8111440GMT2009},
\ref{app:SatJan9154344GMT2010}) along with further material outside
the scope of this study (\ref{app:TueDec29173303GMT2009},
\ref{app:SunFeb1121831CET2009}).

\section{Theoretical model}
\label{sec:MonFeb16112240CET2009}

The Hamiltonian for two coupled qubits reads:
\begin{equation}
  \label{eq:SunDec21163832CET2008}
  H_0=\omega_1\ud{\sigma_1}\sigma_1+\omega_2\ud{\sigma_2}\sigma_2+g(\ud{\sigma_1}\sigma_2+\ud{\sigma_2}\sigma_1)\,,
\end{equation}
where~$\sigma_{1,2}$ are the lowering operators of the qubits, with
bare energies~$\omega_{1,2}$. They are linearly coupled with
strength~$g$. The two modes can be detuned, by a
quantity~$\Delta=\omega_1-\omega_2$, that is small enough,
$\Delta\ll\omega_{1,2}$, so that the rotating wave approximation is
justified. The Hilbert space of the coupled system has dimension four,
with the structure $2\otimes 2$. It can be decomposed in three
subspaces (also called \emph{manifolds}, \emph{rungs}, etc.) with a
fixed number of excitations: the ground state, $\{\ket{0,0}\}$, with
zero excitation, the excited state of each qubit, $\{\ket{1,0},
\ket{0,1}\}$, with one excitation, and the state $\{\ket{1,1}\}$ with
two excitations.

An important point throughout this text is the commutation rules in
Eq.~(\ref{eq:SunDec21163832CET2008}), that are those of two
distinguishable systems, i.e.,
\begin{subequations}
  \label{SunNov15121647GMT2009}
  \begin{align}
    &\sigma_i\sigma_i=\ud{\sigma}_i\ud{\sigma}_i=0\,,\quad i=1,2,\,,\\
    &[\sigma_i,\ud{\sigma}_i]_+=\sigma_i\ud{\sigma}_i+\ud{\sigma}_i\sigma_i=1\,,\quad i=1,2\,,\\
    &[\sigma_i,\ud{\sigma}_j]=\sigma_i\ud{\sigma}_j-\ud{\sigma}_j\sigma_i=0\,,\quad i\neq j\,,\\
    &[\sigma_i,\sigma_j]=\sigma_i\sigma_j-\sigma_j\sigma_i=0\,,\quad i\neq j\,.
  \end{align}
\end{subequations}
Note that two operators from different systems commute.  These
commutation rules for two-level systems is most commonly found in the
literature of the Dicke model~\cite{dicke54a}, that describes a gas of
two-level systems emitting in a common radiation field. In the case of
a fermion gas, this commutation is an approximation, that is made for
the simplicity of the algebra and that is justified for a dilute gas
by the fact that anticommuting operators give the same final physical
results~\cite{shirokov90a}. Indeed, when the wavefunctions of any two
fermions is weakly-overlapping, symmetrized, antisymmetrized and
non-symmetrized results are the same~\cite{messiah_book99a}.

In our case, the two qubits are strongly interacting, which sets our
system apart from the Dicke model and its approximations in many
respects (see appendix~\ref{app:TueDec29173303GMT2009}). For
reference, I will analyze the antisymmetrized case of two interacting
fermions (Fig.~\ref{fig:SatJan9162053GMT2010}(d)) completely. However,
the main object of interest in this text is that of two
\emph{commuting} qubits (like in the Dicke model), corresponding to
the case of two distinguishable (in the quantum sense) qubits
(Fig.~\ref{fig:SatJan9162053GMT2010}(c)).

\begin{figure}[t]
\centering
\includegraphics[width=\linewidth]{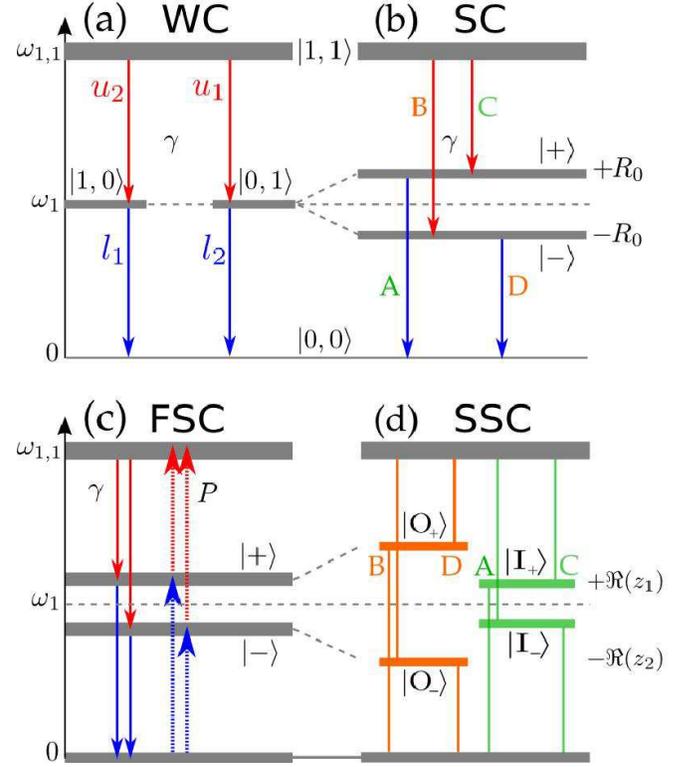}
\caption{(Color online) Energy levels for the two qubits described by
  Hamiltonian~(\ref{eq:SunDec21163832CET2008}). Weak coupling (a) and
  strong coupling (b) in the absence of pump are well defined in terms
  of the bare and dressed states~$\ket{\pm}$, respectively. As shown
  in Sec.~\ref{sec:FriFeb6020814CET2009}, the SC regime gives rise to
  new regions when pump is taken into account: (c) FSC where dressed
  states remain $\ket{\pm}$ , an (d) SSC an MC, where dressed states
  form a new set $\ket{I_\pm}$, $\ket{O_\pm}$. Only SSC, with
  splitting of all the new dressed states, is shown. MC corresponds to
  the case where $\ket{I_\pm}$ have closed (both collapsing on
  $\omega_1$).  The thickness of the levels represents the uncertainty
  in energy due to (a,b) the decay and (c,d) both the pump and the
  decay. The arrows linking the levels due to pump/decay are blue for
  the lower ($l_i$) and red for the upper ($u_i$)
  transitions. Transitions labeled $A$, $C$ (involving $\ket{+}$ or
  $\ket{\mathrm{I_\pm}}$, in green) occur at frequencies determined
  by~$z_1$, while $B$ and $D$ (involving $\ket{-}$ or
  $\ket{\mathrm{O_\pm}}$, in orange) are determined by~$z_2$. The same
  color code is used in the rest of the figures to plot the
  decomposition of the spectra.}
\label{fig:FriJan23211556GMT2009}
\end{figure}

The level structure of Hamiltonian~(\ref{eq:SunDec21163832CET2008}) is
sketched in Fig.~\ref{fig:FriJan23211556GMT2009}(a) (at
resonance). Such a ``diamond-like'' configuration can be mapped to a
four-level system (4LS), i.e., as a single entity, with Hilbert space
structure $1\oplus1\oplus1\oplus1$
(Fig.~\ref{fig:SatJan9162053GMT2010}(e)). This description fits, for
instance, the case of a multi-level atom~\cite{morigi02a} or of a
single quantum dot that can host up to two interacting excitons
(electron-hole pairs) forming a \emph{biexciton}
state~\cite{delvalle10a}. On the other hand, two coupled qubits,
matches the case of two nearby quantum dots directly coupled. The
state with double excitation is then called an \emph{interdot
  biexciton state}~\cite{gerardot05a}.

The Hamiltonian~$H_0$ can be diagonalized in terms of two intermediate
dressed states,~$\ket{+}$ and~$\ket{-}$, as:
\begin{equation}
\label{eq:TueMay5132743GMT2009}
H_0=\omega_{-}\ket{-}\bra{-}+\omega_{+}\ket{+}\bra{+}+\omega_{11}\ket{1,1}\bra{1,1}
\end{equation}
with eigenfrequecies:
\begin{subequations}
  \label{eq:TueMay5124855GMT2009}
  \begin{align}
    &\omega_\pm=\frac{\omega_1+\omega_2}2\pm\mathcal{R}\,,\quad\mathcal{R}=\sqrt{g^2+\left(\frac{\Delta}2\right)^2}\,,\\
    &\omega_{11}=\omega_1+\omega_2\,.\label{eq:FriJan8100155GMT2010}
  \end{align}
\end{subequations}
The diagonalized level structure is sketched in
Fig.~\ref{fig:FriJan23211556GMT2009}(b).  These are the same
eigenfrequecies $\omega_{\pm}$ and Rabi splitting (given
by~$2\mathcal{R}$) than for the dressed states of two coupled harmonic
oscillators up to the first manifold~\cite{laussy09a}. This
equivalence breaks in the manifold with two excitations where the
fermionic nature of the particles reveals and only the state
$\ket{1,1}$ is permitted, as compared to three possible states in the
second manifold of the linear model:
$\{\ket{2,0},\ket{1,1},\ket{0,2}\}$.

The dynamics of dressed states $\ket{\pm}$ and their spectral shape
depend on the amount of decoherence that the dissipative and
excitation processes induce in the system. I will consider an incoming
flow of excitations that populate the two-level systems at rates
$P_1$, $P_2$ and an outgoing flow (given by the inverse lifetime) at
rates $\gamma_1$, $\gamma_2$, respectively. This situation corresponds
to an incoherent continuous pump or injection in the qubit that can be
varied independently from the dissipation. The steady state reached
under the pump and decay corresponds to a statistical mixture of all
possible quantum states and is described by a density
matrix~$\rho$. The master equation of the system has the standard
Liouvillian form~\cite{carmichael_book02a}, with the corresponding
Lindblad terms:
\begin{subequations}
  \label{eq:ME}
\begin{align}
  \frac{d\rho}{dt}=\mathcal{L}\rho=&i[\rho,H]\\
  +&\sum_{i=1,2}\frac{\gamma_i}{2}(2\sigma_i\rho\ud{\sigma}_i-\ud{\sigma}_i\sigma_i\rho-\rho\ud{\sigma}_i\sigma_i)\\
  +&\sum_{i=1,2}\frac{P_i}{2}(2\ud{\sigma}_i\rho\sigma_i-\sigma_i\ud{\sigma}_i\rho-\rho\sigma_i\ud{\sigma}_i)\,.
\end{align}
\end{subequations}
This master equation can be exactly solved given the finite and small
dimension of the Hilbert space (which is only four). Within the same
formalism, we can describe the spontaneous emission from a general
initial state by solving the equations for vanishing pumping.

I will note the transitions between bare states
(cf. Fig.~\ref{fig:FriJan23211556GMT2009}(a)) as:
\begin{subequations}
  \label{eq:ThuOct15135642GMT2009}
  \begin{align}
    u_1=\ket{0,1}\bra{1,1}=\sigma_1 \ud{\sigma_2}\sigma_2 \,,\\
    l_1=\ket{0,0}\bra{1,0}=\sigma_1-u_1 \,, \\
    u_2=\ket{1,0}\bra{1,1}=\ud{\sigma_1}\sigma_1 \sigma_2\,, \\
    l_2=\ket{0,0}\bra{0,1}=\sigma_2-u_2 \,.
  \end{align}
\end{subequations}
(for \emph{upper} and \emph{lower} transition). Note that they can all
be written in terms of the qubit operators (in normal order) since
$\sigma_i=u_i+l_i$ ($i=1,2$). The Lindblad terms of pump and decay in
Eq.~(\ref{eq:ME}) are expressed in terms of their $\sigma$
operators. Rewriting them in terms of the transition operators
$u_{1,2}$, $l_{1,2}$ leads to \emph{cross Lindblad terms} that
entangle $l_1$ and $u_1$, on the one hand, and $l_2$ and $u_2$ on the
other. If the four levels did not correspond to two qubits but to a
single entity (a 4LS), such as atomic levels or a single quantum dot
levels, the Lindblad terms would be written directly in terms of the
transition operators, without these cross Lindblad terms. The
alternative master equation is discussed in
appendix~\ref{app:WedJul8111440GMT2009}
(cf.~Eq.~(\ref{eq:WedJul8112044GMT2009})).

We introduce \emph{fermionic effective broadenings}:
\begin{subequations}
  \label{eq:SunDec21165735CET2008}
  \begin{align}
    &\Gamma_1=\gamma_1+P_1\,,\quad \Gamma_2=\gamma_2+P_2\,,\label{eq:FriJan8110055GMT2010}\\
    &\Gamma_{\pm}=\frac{\Gamma_1\pm\Gamma_2}{4}\,,\quad \gamma_{\pm}=\frac{\gamma_ {1}\pm\gamma_2}{4}\,.\label{eq:FriJan8110105GMT2010}
\end{align}
\end{subequations}

Equation~(\ref{eq:FriJan8110055GMT2010}) is to be compared with
\emph{bosonic effective broadenings}, for boson modes $a$ and $b$
(such as the harmonic oscillators of the linear model):
\begin{equation}
  \label{eq:SunJan10094508GMT2010}
  \tilde\Gamma_a=\gamma_a-P_a\,,\quad\tilde\Gamma_b=\gamma_b-P_b\,.
\end{equation}
A tilde is being used to denote the bosonic character of the
broadening, in the sense that the pumping strength is subtracted to
the decay. Pumping leads to broadening of the line in the fermionic
case and narrowing in the bosonic case, which is a spectral
manifestation of Fermi and Bose statistics.

Another convenient notation for Eqs.~(\ref{eq:SunDec21165735CET2008})
is the expression of pumping and decay rates in terms of $\Gamma$s and
a new parameter $r$ which represents the type of reservoir that the
qubit is in contact with~\cite{briegel93a}:
\begin{subequations}
  \label{eq:TueOct13194019GMT2009}
  \begin{align}
    &\gamma_1=\Gamma_1(1-r_1)\,,\quad \gamma_2=\Gamma_2(1-r_2)\,,\\
    &P_1=\Gamma_1 r_1\,,\quad P_2=\Gamma_2 r_2\, .
\end{align}
\end{subequations}
All possible situations (with $0\leq r_i \leq 1$), from a medium that
only absorbs excitation, $r_i=0$, to one which only provides them,
$r_i=1$, are thus included in a transparent way. For instance, a
thermal bath with temperature different from zero corresponds to
$r_i<1/2$. The effect of the medium on the effective broadening is
contained in $\Gamma_i$.

The \emph{power spectrum} for each qubit, $i=1,2$, is defined as:
\begin{equation}
  \label{eq:TueMay5153242GMT2009}
  s_i(\omega)=\mean{\ud{\sigma}_i(\omega)\sigma_i(\omega)}=\frac{1}{2\pi}\Re\int_0^{\infty}\int_0^{\infty}G_i^{(1)}(t,t+\tau) e^{i\omega\tau}dtd\tau
\end{equation}
where 
\begin{equation}
  \label{eq:TueMay5174910GMT2009}
  G_i^{(1)}(t,t+\tau)=\mean{\ud{\sigma_i}(t)\sigma_i(t+\tau)}\,,
\end{equation}
is the first order auto correlation function. $s(\omega)$ describes
how energy is distributed and is thus of fundamental interest. In a
quantum optical context, this can be observed directly in the optical
emission, but for generality, I will keep the terminology of power
spectrum. For the steady state spectrum, the running time~$t$ is taken
at infinite values (thereby removing one integral). In
appendix~\ref{app:MonJul6000455GMT2009}, we make use of the quantum
regression formula \cite{carmichael_book02a} in its most general form
to compute $G_i^{(1)}$ as well as other two- and one-time correlators
(second order correlation functions are given in
appendix~\ref{app:SunFeb1121831CET2009}). We shall focus on $i=1$ in
the following, without loss of generality:
\begin{equation}
  \label{eq:TueDec23152440CET2008}
  G_1^{(1)}(t,t+\tau)=n_1\sum_{p=A,B,C,D} [L_p(t)+i K_p(t)] e^{-i\omega_p\tau}e^{-\frac{\gamma_p}{2}\tau}\,,
\end{equation}
where $n_1=\int s_1(\omega)d\omega$, the population of qubit~1, is
used to normalize the expression for the spectrum (so that $\int
S_1(\omega)d\omega=1$):
\begin{multline}
  \label{eq:WedDec24161038CET2008}
  S_1(\omega)=\frac1{\pi}\sum_{p\in\{A,B,C,D\}}\Bigg[L_p\frac{\frac{\gamma_p}{2}}{\big(\frac{\gamma_p}{2}\big)^2+(\omega-\omega_p)^2}\\-K_p\frac{\omega-\omega_p}{\big(\frac{\gamma_p}{2}\big)^2+(\omega-\omega_p)^2}\Bigg]\,,
\end{multline}
from Eqs.~(\ref{eq:TueMay5153242GMT2009}) and~(\ref{eq:TueDec23152440CET2008}).

The spectrum is composed of four peaks that I label~$p=A,B,C,D$, each
of them with a Lorentzian (weighted by the coefficient $L_p$) and a
dispersive part (weighted by $K_p$).  The four resonant frequencies
$\omega_p$ and the associated broadenings (full-widths at half
maximum) $\gamma_p$, are intrinsic to the system, as they correspond
to the four possible transitions in the system. The coefficients $K_p$
and $L_p$ (derived in appendix~\ref{app:MonJul6000455GMT2009}) are the
parameters that are specific to the experimental configuration (such
as channel of detection) or regime (steady state under incoherent
pumping or spontaneous emission of an initial state). This form of
$S(\omega)$ is a general feature for the power spectra of coupled
quantum systems~\cite{delvalle09b}.

In the case of uncoupled qubits ($g=0$), the four resonances reduce to
the two bare energies $\omega_1$ and $\omega_2$, broadened by the
effective decay rates~$\Gamma_1$ and $\Gamma_2$, respectively. The
peaks are, in this case, pure Lorentzians. In the opposite case of
very strong coupling ($g\gg\gamma,P$), the spectrum is also well
approximated by Lorentzians, but with the resonances $\omega_p$ now at
the dressed state frequencies~$\omega_\pm$, and broadened by the
average rates~$(\Gamma_1+\Gamma_2)/2$. Lorentzian lineshapes
correspond to the emission of well defined isolated modes of the
system Hamiltonian, weakly affected by other modes. In this case,
$K_p\approx0$. The dispersive contribution becomes non negligible in
the intermediate situations when dissipation and decoherence (or
dephasing, cf.~appendix~\ref{app:TueDec29173303GMT2009}) are of the
order of the direct coupling. The Hamiltonian eigenmodes are then no
longer neatly leading the dynamics and, as a consequence, their broad
emission lines overlap in energy, producing interferences. This is the
regime of interest in our analysis, since new phenomenology appears
for the dressed states and coupling regimes. With this goal in mind, I
devote the next two sections to presenting the analytical expressions
of all the quantities appearing in the spectrum,
Eq.~(\ref{eq:WedDec24161038CET2008}).

\section{Single-time dynamics}
\label{sec:SunFeb1051813CET2009}

We start by analyzing the relevant average quantities needed to
compute the weights $L_p$ and~$K_p$ in the
spectrum~(\ref{eq:WedDec24161038CET2008}), that is, populations and
coherences:
\begin{subequations}
  \begin{align}
    \label{eq:TueMay5193456GMT2009}
    &n_1=\langle\ud{\sigma_1}\sigma_1\rangle\,,\quad n_2=\langle\ud{\sigma_2}\sigma_2\rangle\,\in\mathbb{R}\,,\\
    &n_\mathrm{corr}=\langle\ud{\sigma_1}\sigma_2\rangle\,\in\mathbb{C}\,,\\
    &n_{11}=\langle\ud{\sigma_1}\sigma_1\ud{\sigma_2}\sigma_2\rangle\,\in\mathbb{R}\,.
  \end{align}
\end{subequations}
$n_i$ is the probability that qubit~$i$ is excited. The sum~$n_1+n_2$,
that can go up to two, is the total excitation in the
system. $n_\mathrm{corr}$ is the effective coherence between the
qubits due to the direct coupling. $n_{11}$ is the joint probability
that both qubits are excited. It is also the population of
state~$\ket{1,1}$. If the qubits were uncoupled, we would
have~$n_{11}=n_1n_2$.

It is important here to outline the difference between
\emph{population of a mode} (say,
$n_1=\langle\ud{\sigma_1}\sigma_1\rangle$ for a qubit and
$n_a=\langle\ud{a}a\rangle$ for an harmonic oscillator,
cf.~Fig.~\ref{fig:SatJan9162053GMT2010}(a) and (c)), and
\emph{population of a state} ($\rho_{10}$ and $\tilde\rho_{10}$ for
the state $\ket{1,0}$ with $\ket{1,0}=\ud{a}\ket{0,0}$ and
$\ud{\sigma_1}\ket{0,0}$, respectively). The population of the
intermediate state $\ket{1,0}$ (resp.~$\ket{0,1}$) is given
by~$n_1-n_{11}$ (resp.~$n_2-n_{11}$). This is also the probability of
having only one of the qubits excited. The population of the ground
state is given by $1-n_1-n_2+2n_{11}$.

In the spontaneous emission case, $n_1$, $n_2$ and~$n_\mathrm{corr}$
have the same solutions (depending only on the initial condition
$n_1^0$, $n_2^0$ and $n_\mathrm{corr}^0$) than their counterpart for
the two coupled linear oscillators, $n_a$, $n_b$ and~$n_\mathrm{ab}$.
However, the two models differ for $n_{11}$: in the case of coupled
qubits, it decays from its initial value,
$n_{11}(t)=e^{-4\gamma_+t}n_{11}^0$, whereas in the linear model, the
population $\tilde\rho_{11}$ oscillates as a result of the exchange
with the other states that are available in the second manifold
($\ket{2,0}$ and $\ket{0,2}$).  Although populations of the modes have
the same dynamics both in the two harmonic oscillators and the two
coupled qubits, the underlying populations of their states do not. For
instance, the decay from the initial condition~$\ket{1,1}$ leads to
$n_1=n_{10}+n_{11}=\rho_{10}+\rho_{11}$ for the qubit, while for the
harmonic oscillators, one has,
$n_a=\mean{\ud{a}a}=\tilde\rho_{10}+\tilde\rho_{11}+2\tilde\rho_{20}$. Since
$n_1=n_a$ (and also $n_2=n_b$) the states are differently populated
($\rho_{10}\neq\tilde\rho_{10}$, Etc.).

In the steady state, all these mean values can be written in terms of effective
pump and decay parameters, as in the two harmonic oscillators, but now
following fermionic statistics $(i=1,2)$:
\begin{subequations}
  \label{eq:SunDec21171436CET2008}
  \begin{align}
    &n_i^\mathrm{SS}=\frac{P_i^\mathrm{eff}}{\gamma_i^\mathrm{eff}+P_i^\mathrm{eff}}\,,\\
    &\gamma_i^\mathrm{eff}=\gamma_i+\frac{\gamma_1+\gamma_2}{\Gamma_1+\Gamma_2}Q_i\,,\quad P_i^\mathrm{eff}=P_i+\frac{P_1+P_2}{\Gamma_1+\Gamma_2}Q_i\,,\\
    &n_\mathrm{corr}^\mathrm{SS}=\frac{g}{\Delta-2i\Gamma_+}(n_1-n_2)\,,
\end{align}
\end{subequations}
with the corresponding generalized Purcell rates
\begin{equation}
  Q_1=\frac{4(g^\mathrm{eff})^2}{\Gamma_2}\,,\quad Q_2=\frac{4(g^\mathrm{eff})^2}{\Gamma_1}\,,
\end{equation}
and the effective coupling strength
\begin{equation}
  \label{eq:ThuJul9102119GMT2009}
  g^\mathrm{eff}=\frac{g}{\sqrt{1+\Big(\frac{\Delta/2}{\Gamma_+}
      \Big)^2}}\,.
\end{equation}
Finally, $n_{11}$ takes a simple intuitive form in the steady state,
\begin{equation}
  \label{eq:SunDec21173823CET2008}
  n_{11}^\mathrm{SS}=\frac{n_1^\mathrm{SS}P_2+n_2^\mathrm{SS}P_1}{\Gamma_1+\Gamma_2}\leq n_1^\mathrm{SS}n_2^\mathrm{SS}\,.
\end{equation}
This should be contrasted with the counterpart of $n_{11}^\mathrm{SS}$
for two bosonic modes ($a$ and $b$), for which:
\begin{equation}
  \label{eq:FriFeb13121116CET2009}
  \langle\ud{a}a\ud{b}b\rangle^\mathrm{SS}=2n_a^\mathrm{SS}n_b^\mathrm{SS}-\frac{n_a^\mathrm{SS}P_b+n_b^\mathrm{SS}P_a}{\Gamma_a+\Gamma_b}\geq
  n_a^\mathrm{SS}n_b^\mathrm{SS}\,.
\end{equation}
This is an interesting manifestation of the symmetry/antisymmetry of
the wavefunction for two bosons/fermions, that is known to produce
such an attractive/repulsive character for the correlators. Here we
see that quantum (or correlated) averages $\langle\hat n_1\hat
n_2\rangle$ (with~$\hat n_i$ the number operator) are higher/smaller
than classical (uncorrelated) averages $\langle\hat n_1\rangle\langle
\hat n_2\rangle$, depending on whether they are of a boson or fermion
character, respectively. This provides a neat picture of
bunching/antibunching from excitations of different modes that are
otherwise of the same character.

In the most general case, with pump and decay, before the steady state
is reached, also the transient dynamics of the mean values $n_1$,
$n_2$ and $n_\mathrm{corr}$ for the coupled qubits maps to the
corresponding averages $n_a$, $n_b$ and $n_{ab}$ of coupled harmonic
oscillators, with only $\tilde \Gamma \rightarrow \Gamma$.  We can
conclude then, that the single-time dynamics of the qubit is ruled by
a (half) Rabi frequency of the same form than in the boson
case~\cite{laussy09a}:
\begin{equation}
  \label{eq:ThuApr3185957CEST2008}
  R^\mathrm{1TD}=\sqrt{g^2-(\Gamma_-+i\Delta/2)^2}\,,
\end{equation}
only with Fermion-like effective broadenings,
Eqs.~(\ref{eq:SunDec21165735CET2008}). Since, in contrast to the boson
case where there is only one Rabi parameter, another expression will
arise in the two-time dynamics for coupled qubits, I will refer to
Eq.~(\ref{eq:ThuApr3185957CEST2008}) as the \emph{single-time
  dynamics (half) Rabi frequency}.

We conclude this section by noting that the magnitudes studied up to
now are independent on having two qubits, two identical fermions or a
4LS. The cross terms appearing in the master equation for the first
case, due to the correlations induced by the incoherent processes, do
not affect the steady state populations, as is shown in
appendix~\ref{app:WedJul8111440GMT2009}.

\section{Power spectra}
\label{sec:WedJul29174809GMT2009}

\begin{figure}[tb]
\centering
\includegraphics[width=\linewidth]{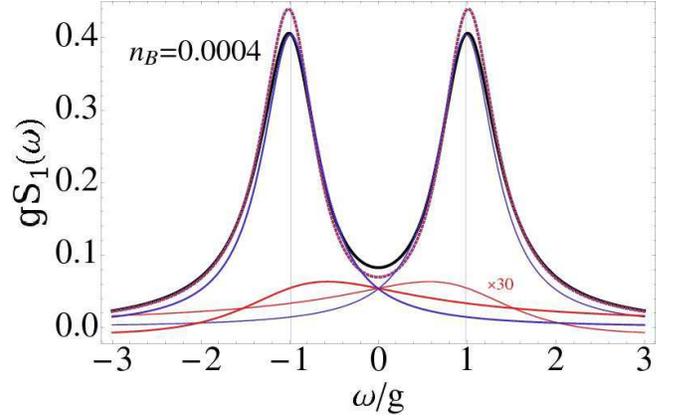}
\caption{(Color online) Power spectrum, $S_1(\omega)$, from a qubit
  (thick solid black line) in the SC regime ($\gamma_1=g$ and
  $\gamma_2=g/2$) for the steady state under vanishing pump
  ($P_1=0.02g$ and $P_2=0.01g$). The spectra is composed of four peaks
  arising from the lower and upper transitions (blue and red thin
  lines, respectively). In dashed purple, the linear model spectrum
  for comparison.}
\label{fig:MonJan26012453GMT2009}
\end{figure}

\subsection{Two qubits}
\label{sec:SunJan10204219GMT2010}

The general expressions for the spectrum and two-time correlators of
two coupled qubits admit analytic solutions at resonance in the steady
state of an incoherent continuous pump. From now on, we will refer
always to this situation and, therefore, I will drop the steady state
label in the notation.

The four coefficients $L_p+iK_p$ appearing in
Eq.~(\ref{eq:TueDec23152440CET2008}), now defined in the steady state,
read: (cf.~appendix~\ref{app:MonJul6000455GMT2009})
\begin{subequations}
  \label{eq:TueDec23121254CET2008}
  \begin{align}
    L_A+iK_A=&\frac{1}{16 R z_1}\Big\{2(2z_1+i\Gamma_2)(R-i\Gamma_-)+a_1+a_2\frac{n_2}{n_1}\nonumber\\
    &+2g\Big[-\frac{P_1}{\Gamma_+}(2z_1+i\Gamma_2)+2(R+z_1+i\Gamma_+)\Big]\frac{n_\mathrm{corr}}{n_1}\Big\}\,,\\
    L_B+iK_B=&\frac{1}{16 R z_2}\Big\{2(2z_2+i\Gamma_2)(R+i\Gamma_-)-a_1-a_2\frac{n_2}{n_1}\nonumber\\
    &+2g\Big[\frac{P_1}{\Gamma_+}(2z_2+i\Gamma_2)+2(R-z_2-i\Gamma_+)\Big]\frac{n_\mathrm{corr}}{n_1}\Big\}\,,\\
    L_C+iK_C=&\frac{1}{16 R z_1}\Big\{2(2z_1-i\Gamma_2)(R-i\Gamma_-)-a_1-a_2\frac{n_2}{n_1}\nonumber\\
    &+2g\Big[-\frac{P_1}{\Gamma_+}(2z_1-i\Gamma_2)-2(R-z_1+i\Gamma_+)\Big]\frac{n_\mathrm{corr}}{n_1}\Big\}\,,\\
    L_D+iK_D=&\frac{1}{16 R z_2}\Big\{2(2z_2-i\Gamma_2)(R+i\Gamma_-)+a_1+a_2\frac{n_2}{n_1}\nonumber\\
    &+2g\Big[\frac{P_1}{\Gamma_+}(2z_2-i\Gamma_2)-2(R+z_2-i\Gamma_+)\Big]\frac{n_\mathrm{corr}}{n_1}\Big\}\,.
\end{align}
\end{subequations}
They are defined in terms of the parameters
\begin{equation}
  \label{eq:ThuJan22124913GMT2009}
  a_1=\frac{g^2}{\Gamma_+^2}[4\Gamma_+^2+2P_1(P_2-2\Gamma_+)-P_2\Gamma_1]\,,\quad a_2=\frac{g^2}{\Gamma_+^2}P_1(P_1-\gamma_1)\,,
\end{equation}
and the corresponding frequencies and decay rates, that also appear
explicitly in Eq.~(\ref{eq:TueDec23152440CET2008}):
\begin{subequations}
  \label{eq:TueDec23152835CET2008}
  \begin{align}
    &\frac{\gamma_A}{2}+i\omega_A=2\Gamma_++iz_1\,,\quad
    &\frac{\gamma_B}{2}+i\omega_B=2\Gamma_++iz_2\,,\\
    &\frac{\gamma_C}{2}+i\omega_C=2\Gamma_+-iz_1\,,\quad
    &\frac{\gamma_D}{2}+i\omega_D=2\Gamma_+-iz_2\,.
  \end{align}
\end{subequations}

They all depend on two complex parameters, $z_1$ and $z_2$:
\begin{equation}
  \label{eq:TueDec23152618CET2008}
  z_{1,2}=\sqrt{(D^\mathrm{s}g)^2+(i\Gamma_+\pm R)^2}\,.
\end{equation}
The \emph{degree of symmetry}, $D^\mathrm{s}$, is a real dimensionless
quantity, between $0$ and $\sqrt{2}$, given by
\begin{equation}
  \label{eq:FriJan30193148CET2009}
  D^\mathrm{s}=\frac{\sqrt{(\gamma_1P_2+\gamma_2P_1)/2}}{\Gamma_+}\,.
\end{equation}
This quantity is proper to the coupled qubits case and its physical
meaning will be clarified later. Its value is linked to the symmetry
between the different parameters. For instance, $D^\mathrm{s}=1$ when
all parameters are equal to each other,
$\gamma_1=\gamma_2=P_1=P_2$. On the other hand, $D^\mathrm{s}=0$ if
one of the parameters (any of them) is much larger than the others. It
leads to a renormalized coupling strength
\begin{equation}
  \label{eq:uniqname}
  G=D^\mathrm{s}g\,,
\end{equation}
that reaches a maximum when the parameters are such that
$D^\mathrm{s}=\sqrt{2}$. Such an enhancement, by~$\sqrt{2}$, is
related to the cooperative behavior of two coupled modes, similarly to
the superradiance of two atoms in the Dicke model or the
renormalization with the mean number of photons in the Jaynes-Cummings
Model.

The last and most important parameter appearing in the previous
expressions is a Rabi frequency for the two-time dynamics, that for
coupled qubits differs from its counterpart for single-time dynamics
(cf.~Eq.~(\ref{eq:ThuApr3185957CEST2008})):
\begin{equation}
  \label{eq:TueDec23154618CET2008}
  R=\sqrt{g^2-(D^\mathrm{s} g)^2-\Gamma_-^2}\,.
\end{equation}
This is the true analog of the (half) Rabi frequency of the linear
model since this value, not its single-time counterpart, determines
strong or weak coupling (emergence of dressed states). At vanishing
pump, the renormalized coupling~$G$ converges to~$g$, and both~$R$
and~$R^\mathrm{1TD}$ converge to the standard expression for the
(half) Rabi splitting~\cite{laussy09a}:
\begin{equation}
  \label{eq:ThuJul9112038GMT2009}
  R_0=\sqrt{g^2-\gamma_-^2}\,.
\end{equation}

The normalized power spectrum of qubit~$1$ follows from
Eq.~(\ref{eq:WedDec24161038CET2008}) with the coefficients we have
obtained. The positions and broadenings of the four peaks are given
respectively by the real and imaginary parts of $z_1$ and $z_2$. Their
expressions remain valid in the spontaneous emission case by setting
the pumping rates to zero.

Figure~\ref{fig:MonJan26012453GMT2009} is an example of the spectrum
$S_1(\omega)$ (in solid black) and its decomposition in four peaks
(thin blue and red). The split positions of the four peaks, which
indicate the system is in the SC regime, are marked with two vertical
blue lines. The two peaks that correspond to the lower manifold
transitions, $A$ and $D$ in Fig.~\ref{fig:FriJan23211556GMT2009}(b),
appear with a thin blue line. Upper transitions, $B$ and $C$ in
Fig.~\ref{fig:FriJan23211556GMT2009}(b), appear with a thin red line.
All resonances are at the same positions but the stronger dispersive
part of upper transitions leads to a shift of their maximum.  The
upper transitions are much weaker in intensity (magnified $\times 30$
to be visible) due to the small pump.  The double excitation of the
system is very unlikely ($n_{11}=0.0004$).  The lineshape is therefore
close to that of two coupled harmonic oscillators, plotted with a
dashed purple line for comparison. The system is in the \emph{linear
  regime} where all models of Fig.~\ref{fig:SatJan9162053GMT2010} for
two coupled modes converge. In the following sections, we will see how
the lineshapes change when entering the nonlinear regime.

\subsection{Two fermions}
\label{sec:FriNov20155955WET2009}

As noted before, although the expressions for the single-time dynamics
(populations, coherence, Etc.) for two coupled fermions (anticommuting
operators for modes $1$ and $2$) are the same than for the two coupled
qubits, their power spectra are different.  As compared to the coupled
qubits, the coupled fermions spectrum assumes a simple and fundamental
form, closely related to that of the two harmonic oscillators: the
formal expression is the same, differing only in the parameters
(effective broadenings, populations, Etc.).  In particular, only one
Rabi parameter, the single-time Rabi frequency, $R^\mathrm{1TD}$,
determines both the single- and two-time dynamics. The two-fermions
spectra are thus obtained by simply substituting the fermionic
parameters (Eq.~\ref{eq:SunDec21165735CET2008}) in the expression of
the linear model~\cite{laussy09a}.

This simplicity and likeliness to the linear model stems from the
fundamental nature of the problem: two identical (indistinguishable)
particles coupled linearly, obeying fully their quantum statistics.
The two coupled qubits (or the Jaynes-Cummings
model~\cite{delvalle09a}), by mixing different types of particles
(distinguishable modes) and therefore breaking commutation rules,
result in the more complex description and richer dynamics presented
in the previous section.

\subsection{four-level system (4LS)}
\label{sec:MoNov23201725WET2009}

Also in the four-level system (with no cross Lindblad terms), the
expressions for the single-time dynamics (populations, coherence,
Etc.) are the same than for the linear model, and here also their
power spectra are different. The parameters for the 4LS spectra are of
a bosonic character, cf.~Eq.~(\ref{eq:SunJan10094508GMT2010}):
\begin{subequations}
  \label{eq:WedAug5130854GMT2009}
  \begin{align}
    &\tilde\Gamma_{1}=\gamma_1-P_1\,,\quad \tilde\Gamma_2=\gamma_2-P_2\,,\\
    &\tilde\Gamma_{\pm}=\frac{\tilde\Gamma_1\pm\tilde\Gamma_2}{4}\,,\\
    &\tilde R=\sqrt{g^2-\tilde\Gamma_-^2}\,.
\end{align}
\end{subequations}
The relevant parameters that characterize the coupling simplify to:
\begin{subequations}
  \label{eq:WedAug5133404GMT2009}
  \begin{align}
    &R=\frac{\tilde\Gamma_+}{\Gamma_+}\tilde R\,,\\
    &z_{1,2}=\tilde R\pm i\tilde\Gamma_+\,,
  \end{align}
\end{subequations}
recovering the conventional strong coupling criterion based on one
parameter only, the bosonic (half) Rabi frequency $\tilde R$. The
resulting spectral structure then consists of two pairs of peaks
sitting at $\pm\Re{(\tilde R)}$ with:
\begin{subequations}
  \label{eq:WedAug5140534GMT2009}
  \begin{align}
    &\frac{\gamma_A}{2}+i\omega_A=\frac{3(P_1+P_2)+\gamma_1+\gamma_2}{4}+i\tilde R\,,\\
    &\frac{\gamma_B}{2}+i\omega_B=\frac{3(\gamma_1+\gamma_2)+P_1+P_2}{4}+i\tilde R\,,\\
    &\frac{\gamma_C}{2}+i\omega_C=\frac{3(\gamma_1+\gamma_2)+P_1+P_2}{4}-i\tilde R\,,\\
    &\frac{\gamma_D}{2}+i\omega_D=\frac{3(P_1+P_2)+\gamma_1+\gamma_2}{4}-i\tilde R\,.
  \end{align}
\end{subequations}
Note that $\gamma_p$ are always positive for any combination of the
parameters, in contrast with those of two bosonic modes, where the
system can diverge. Therefore, the values of pump and decay rates here
are not limited, always leading to a physical steady state.

\section{Strong and Weak coupling regimes}
\label{sec:MonFeb16112450CET2009}

The standard criterion for strong coupling (SC) is based on the
splitting at resonance of the bare states into dressed states. This
manifests in the appearance of $\tau$-oscillations in the two-time
correlators and a splitting of the peaks that compose their spectrum.

In a naive approach to the problem of defining strong-coupling in a
system other than the linear model, one could think that the condition
for SC is $\Re({R^\mathrm{1TD}})\neq 0$ (at resonance), leading to the
familiar inequality, $g>|\Gamma_-|$. However, this is not the case
whenever pump and decay are both taken into account. Instead, one must
find the condition for a splitting between the new eigenstates, that
is, the two pairs of peaks forming the spectrum. The peaks are
positioned symmetrically in two pairs about the origin at
$\omega_p=\pm\Re{(z_{1,2})}$ and, therefore,
\begin{equation}
\label{eq:SatJan24191208GMT2009}
\Re{(z_1)}\neq 0\quad\mathrm{or}\quad\Re{(z_2)}\neq 0
\end{equation}
is the mathematical condition for SC in this system. Given that there
are two different parameters $z_1$ and $z_2$ on which the condition
relies, the SC/WC distinction must be extended to cover new
possibilities. Thus, instead of only one relevant parameter,
$\Gamma_-/g$, as was the case in the linear model, SC between two
qubits is determined by three parameters:
\begin{equation}
  \label{eq:SatJan31220509CET2009}
  \Gamma_-/g\,,\quad\,\Gamma_+/g\quad\text{and}\quad\,D^\mathrm{s}\,.
\end{equation}

This gives rise to the situations listed in
Table~\ref{tab:SatJan9121355GMT2010}, that are discussed in the
following sections.

\begin{table}[ht]
  \centering
  \begin{tabular}{| c | c | c | c | c | l |}
    \hline
    $R$ & $\Re(z_1)$ & $\Re(z_2)$ & Acronym & Type of coupling \\
    \hline
    \hline
    $|R|$ & $\neq0$ & $\neq0$  & FSC & First order   Strong Coupling \\
    $i|R|$ & $\neq0$ & $\neq0$  & SSC & Second order Strong Coupling \\
    $i|R|$ & 0 & $\neq0$  & MC & Mixed Coupling \\
    $i|R|$ & 0 & 0 & WC & Weak Coupling \\
    \hline
  \end{tabular}
  \caption{Type and nomenclature of coupling for two coupled qubits. Beyond the usual weak coupling (WC) and strong coupling (here denoted FSC) encountered in the linear model, the system exhibits two new regions: Mixed Coupling (coexistence of weak and strong coupling) and Second order Strong Coupling (with two different splittings of two pairs of dressed states).}
  \label{tab:SatJan9121355GMT2010}
\end{table}

\subsection{Vanishing pump and spontaneous emission}
\label{sec:SunJan10204807GMT2010}

In the case of vanishing pump, that corresponds as well to spontaneous
emission, the standard SC and WC hold. In this limit, we recover the
familiar expression for the half Rabi frequency
$R,R^\mathrm{1TD}\rightarrow R_0$. The parameters simplify to
$z_{1,2}\rightarrow\sqrt{(R_0\pm i\gamma_+)^2}=R_0\pm
i\gamma_+$~\footnote{When taking square roots, we always follow the
  prescription $\sqrt{c^2}=c$, for $c\in\mathbb{C}$. Considering this
  solution is enough because in any case all quantities depend on both
  $\pm \sqrt{c^2}$.}.

The positions and broadenings of the four peaks are:
\begin{subequations}
  \label{eq:ThuDec25175033CET2008}
  \begin{align}
    &\frac{\gamma_A}{2}+i\omega_A=\gamma_++iR_0\,,\quad&\frac{\gamma_B}{2}+i\omega_B=3\gamma_++iR_0\,,\\
    &\frac{\gamma_C}{2}+i\omega_C=3\gamma_+-iR_0\,,\quad&\frac{\gamma_D}{2}+i\omega_D=\gamma_+-iR_0\,.
  \end{align}
\end{subequations}
From here, the associated condition for SC reduces to $R_0$ being
real, or more explicitly $g>|\gamma_-|$, as in the linear model at
vanishing pump [see Fig.~\ref{fig:FriJan23211556GMT2009}(b)]. In SC,
the two pairs of peaks $p=A,D$ and $p=B,C$ sit on the same frequencies
although they have different broadenings. Excited states have shorter
lifetime, since each excitation can decay. From
Eqs.~(\ref{eq:ThuDec25175033CET2008}), the two dressed states undergo
the transition into weak coupling (WC) simultaneously [as in
Fig.~\ref{fig:FriJan23211556GMT2009}(a)]. In WC, $R_0\rightarrow
i|R_0|$ and both parameters $z_{1,2}$ become imaginary, giving
\begin{subequations}
  \label{eq:MonJan26115601GMT2009}
  \begin{align}
    &\omega_p=0\,,\quad p=A,B,C,D\,,\\
    &\frac{\gamma_A}{2}=\gamma_+-|R_0|\,,\quad\frac{\gamma_B}{2}=3\gamma_+-|R_0|\,,\\
    &\frac{\gamma_C}{2}=3\gamma_++|R_0|\,,\quad\frac{\gamma_D}{2}=\gamma_++|R_0|\,,
  \end{align}
\end{subequations}
with $\gamma_p\geq 0$, since $\gamma_+\geq|R_0|$ in this regime. The
four peaks collapse into four Lorentzians at the origin, all differing
in their broadenings.

As a result of the two pairs of peaks sitting always on the same two
(or one) frequencies, the final spectra can only be either a single
peak or a doublet, both shapes being possible in SC or WC regimes (as
in the linear model and for the same reasons~\cite{laussy09a}).  An
intuitive derivation and interpretation of these results is given in
appendix~\ref{app:SatJan9154344GMT2010}, based on the so-called
\emph{manifold picture}~\cite{delvalle_book09a}, which consists in
considering transitions between eigenstates of a non-hermitian
Hamiltonian, with energies broadened by the imaginary part.

In the steady state case but in the limit of vanishing pump (the
linear regime), only the vacuum and first manifold are populated. The
spectra in this limit converge with the linear model and also it can
be analyzed in terms of manifolds by straightforward extension. The
spectrum in Fig.~(\ref{fig:MonJan26012453GMT2009}) is an example of SC
for vanishing pump as we can see from the fact that the lower
transition peaks, in blue, dominate over the broader and weak upper
peaks, in red. In this case, the splitting of the dressed modes gives
rise to a splitting in the final spectrum (in black). In what follows,
we take this SC configuration ($\gamma_1=g$ and $\gamma_2=g/2$) as a
starting point to explore the effect of a non-negligible incoherent
continuous pump.

\subsection{Non-negligible pump}
\label{sec:FriFeb6020814CET2009}

\begin{figure}[t]
\centering
\includegraphics[width=0.9\linewidth]{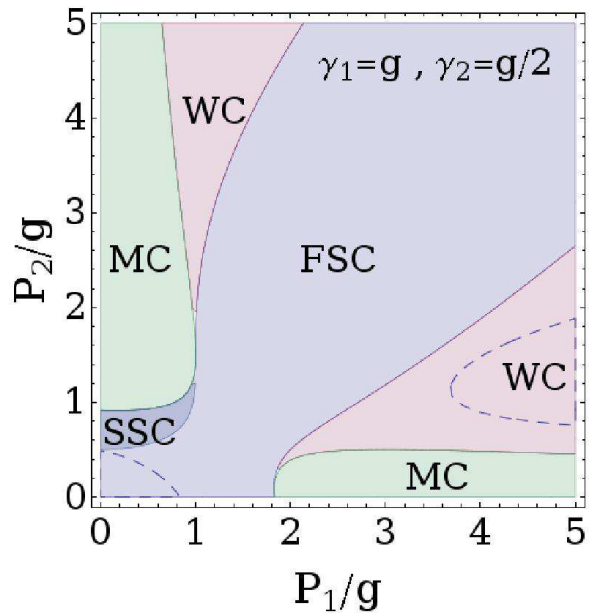}
\caption{(Color online) Phase space of the steady state Strong/Weak
  Coupling regimes as a function of pump for $\gamma_1=g$ and
  $\gamma_2=g/2$. In Strong Coupling (SC, blue), one can distinguish
  two regions, First order (FSC, light blue) and Second order (SSC,
  dark blue) Strong Coupling. Weak Coupling (WC) is in purple and
  Mixed Coupling (MC) in green. The dashed blue lines enclose the two
  regions where two peaks can be resolved in the power spectrum of the
  first qubit, $S_1(\omega)$. One falls in SC and the other in WC.}
\label{fig:SatJan24165755GMT2009}
\end{figure}

When pump is taken into account, all the types of coupling listed in
Table~\ref{tab:SatJan9121355GMT2010} are accessible. These are plotted
in Fig.~\ref{fig:SatJan24165755GMT2009} as a function of the pumping
rates.  The starting point is the Rabi frequency~$R$,
Eq.~(\ref{eq:TueDec23154618CET2008}), that is either real or pure
imaginary.

First, let us consider the case where:
\begin{equation}
\label{eq:SunJan25110437GMT2009}
R=|R|\quad\Leftrightarrow \quad G^2<g^2-\Gamma_-^2\,,
\end{equation}
from which follows that $z_1=z_2^*$, and therefore
$\Re{(z_1)}=\Re{(z_2)}$. This is the most standard situation that we
already found in the absence of pumping. It is sketched in
Fig.~\ref{fig:FriJan23211556GMT2009}(c). To distinguish it from the
other types of coupling to be discussed shortly, I will from now on
call it \emph{First order Strong Coupling} (FSC). Note that
condition~(\ref{eq:SunJan25110437GMT2009}) can only be satisfied if
$D^\mathrm{s}<1$, therefore, when the renormalization of the coupling
through the interplay of pump and decay is detrimental (since
Eq~(\ref{eq:SunJan25110437GMT2009}) implies that $G<g$). It is not
possible to reach the optimum effective coupling and maximum splitting
of the spectral lines given by
$2\sqrt{2}g$. Eq.~(\ref{eq:SunJan25110437GMT2009}) leads to the
following explicit condition for $g$:
\begin{equation}
  \label{eq:SunFeb8004851CET2009}
  g>\frac{|\Gamma_-|}{\sqrt{|1-(D^\mathrm{s})^2|}}\,.
\end{equation} 

If $R\neq 0 $, then, also $\Re{(z_1)}=\Re{(z_2)}\neq 0$. FSC includes
the standard SC regime in the absence of pump ($g>|\Gamma_-|$ that is
implied by Eq.~(\ref{eq:SunFeb8004851CET2009})). It is the most
extended region in Fig.~\ref{fig:SatJan24165755GMT2009}, colored in
light blue. In this case, the spectrum of emission follows the
expected pattern: two pairs of peaks, $A$, $D$ and $B$, $C$, are
placed one on top of each other, although they are differently
broadened [see the spectra in Fig.~\ref{fig:MonJan26012453GMT2009}].

\begin{figure}[t]
\centering
\includegraphics[width=\linewidth]{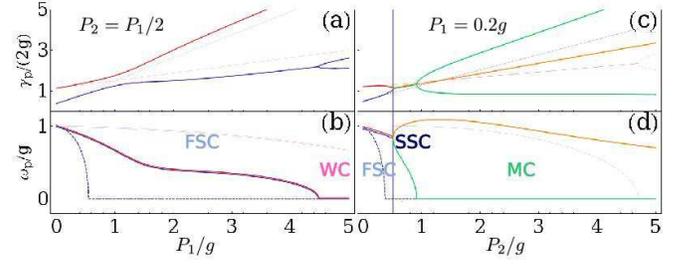}
\caption{(Color online) Broadenings (a), (c), and positions (b), (d)
  of the lines that compose the spectra as a function of pump for the
  decay parameters $\gamma_1=g$ and $\gamma_2=g/2$. In the plots of
  the first column, the pump $P_1$ varies with $P_2=P_1/2$, moving
  upwards in the phase space of
  Fig~\ref{fig:SatJan24165755GMT2009}. The vertical guideline shows
  the crossing from FSC to WC. In the plots of the second column, the
  pump $P_2$ varies with $P_1=0.2g$, moving in diagonal in the phase
  space of Fig~\ref{fig:SatJan24165755GMT2009}. The vertical
  guidelines show the crossing from FSC to SSC and finally to MC. The
  dashed blue line represents the splitting as it is resolved in the
  final spectrum~$S_1(\omega)$. The color code (red-blue and
  orange-green) corresponds to that of the transitions in
  Fig.~\ref{fig:FriJan23211556GMT2009}.}
\label{fig:SatJan24165822GMT2009}
\end{figure}

In Fig.~\ref{fig:SatJan24165822GMT2009}(a) and (b) we track the
broadenings and positions of the four peaks ($A$ and $D$ in blue and
$B$ and $C$ in red) as a function of pump, through the SC region of
Fig~\ref{fig:SatJan24165755GMT2009}, on a diagonal line defined
by~$P_2=P_1/2$. Following them from vanishing pump, where the manifold
picture is exact, the four peaks can be easily associated with the
lower and upper transitions of
Fig.~\ref{fig:FriJan23211556GMT2009}(b), and that is why we keep the
same color code and notation. Dressed states $\ket{\pm}$, close to the
Hamiltonian ones, can still be defined in the system, but with the
modified frequencies~$\omega_1\pm\Re{(z_1)}$, both affected equally by
decoherence.

By construction, the resulting spectra in this regime can only be a
doublet or a single peak, depending on the magnitude of the broadening
of the peaks (that always increases with pump and decay) against the
splitting of the lines (that always decreases). As in the limit of
vanishing pump, observing a doublet in the spectra does not imply
splitting of the dressed states (and thus, SC)~\cite{laussy08a}, but
here the tendency is always the same: the lower the pump and the
decay, the better the resolution of the splitting.

Second, let us consider the situation of the Rabi frequency being
imaginary:
\begin{equation}
\label{eq:MonJan26185254GMT2009}
R=i|R|\quad\Leftrightarrow \quad G^2>g^2-\Gamma_-^2\,.
\end{equation}
This results in three possibilities, listed in
Table~\ref{tab:SatJan9121355GMT2010} (WC, SSC and MC), that constitute
the three remaining regions delimited in
Fig.~\ref{fig:SatJan24165755GMT2009}. In what follows we find the
specificities of each of these three regimes.

The \emph{Weak Coupling} regime (WC, in purple) is characterized by
\begin{equation}
\label{eq:MonJan26190025GMT2009}
z_1=i|z_1|,\quad z_2=i|z_2|,\quad z_1\neq z_2\quad\Leftrightarrow \quad G<|\Gamma_+-|R||
\end{equation}
and therefore $\Re{(z_1)}=\Re{(z_2)}=0$. Note that
condition~(\ref{eq:MonJan26190025GMT2009}) is not analytical in terms
of the relevant parameters~(\ref{eq:SatJan31220509CET2009}). In WC,
the four peaks are placed at the origin with four different
broadenings. The dressed states have collapsed in energy
to $\omega_1$.

Up to here, we have remained within the SC and WC regions already
known from the linear model. We now consider the two new regions of
SC, proper to the coupled qubits, that I call SSC and MC,
respectively:

\paragraph*{SSC:}

When both parameters $z_{1,2}$ are real, then:
\begin{equation}
  \label{eq:MonJan26211714GMT2009}
  z_1=|z_1|,\quad z_2=|z_2|,\quad z_1< z_2\quad\Leftrightarrow \quad G>|\Gamma_++|R||\,.
\end{equation}
We refer to it as \emph{Second order Strong Coupling} regime (SSC,
colored in dark blue in the phase space). Here, the broadenings of the
four peaks are equal, $\gamma_p/2=2\Gamma_+$, but the positions of the
pairs of peaks are different, $\omega_{A,C}=\pm |z_1|$ and
$\omega_{B,D}=\pm |z_2|$. The reason is that the bare energies of the
modes undergo a \emph{second order anticrossing} induced by the
interplay between coupling, pump and decay. The energies of the
dressed states are affected differently by decoherence, up to the
point where we may picture the physics in terms of a new type of
eigenstates.  The association of the $A$ and $D$ (with $\omega_{A,D}$)
as the peaks corresponding to lower transitions and $B$ and $C$ (with
$\omega_{B,C}$) to upper transitions is completely arbitrary in this
region, given that the broadenings of the peaks, which led us to such
association in FSC, are now equal. This implies that, rather than two
dressed states, $\ket{-}$ and $\ket{+}$
(Fig.~\ref{fig:FriJan23211556GMT2009}(b)) as in the conventional
strong coupling (FSC), the system now exhibits four dressed states:
$\ket{I_-}$, $\ket{I_+}$, $\ket{O_-}$ and $\ket{O_+}$. They are
plotted in Fig.~\ref{fig:FriJan23211556GMT2009}(d):
$\ket{\mathrm{I_\pm}}$ (resp. $\ket{\mathrm{O_\pm}}$) have energies
split at $\pm|z_1|$, giving rise to the \emph{inner peaks}, in green
(resp. $\pm|z_2|$, giving rise to the \emph{outer peaks}, in
orange). The physical origin of this remarkable departure from the
conventional strong coupling picture will be discussed in
section~\ref{sec:WedJul29173348GMT2009}.

We can see how peak broadenings and positions change when going from
FSC to SSC in Fig.~\ref{fig:SatJan24165822GMT2009}(c) and (d). In this
case, we track the peaks by varying $P_2$ for a fixed $P_1$, moving
upwards in the phase space. The first vertical guideline marks the
border between the two kinds of SC, with the opening of a ``bubble''
for the positions $\omega_A$ and $\omega_B$ (that were equal in the
FSC region), and the convergence of all the broadenings.  In
principle, one can expect that quadruplets and triplets may form out
of the four peaks. However, the broadenings and contributions of the
dispersive parts (given by~$K_p$) are too large to let any fine
splitting emerge clearly. The spectra in this region reduce to
singlets and doublets. However, we show in
Sec.~\ref{sec:SunFeb1004214CET2009} through some examples that they
may be distorted, doubtlessly reflecting the multiplet structure.

\paragraph*{MC:}

When $z_1$ is imaginary and $z_2$ real, or equivalently,
\begin{equation}
  \label{eq:MonJan26211714GMT20092}
  z_1=i|z_1|\,,\quad z_2=|z_2|\quad\Leftrightarrow \quad |\Gamma_+-|R||<G<|\Gamma_++|R||\,,
\end{equation}
we enter the last new region in Fig.~\ref{fig:SatJan24165755GMT2009}.
This is a \emph{Mixed Coupling} regime (MC, colored in green in the
phase space) where the two inner peaks, A and C---as well as the
reconstructed eigenstate $\ket{\mathrm{I}_\pm}$---have collapsed at
the origin, like in WC. However, the two outer peaks, B and D---as
well as $\ket{\mathrm{O}_\pm}$---are still split. As in SSC, the
broadening of the peaks does not allow for a distinction between upper
of lower resonances. The collapsed resonance is in this case at the
bare energy $\omega_1=0$ because that is the total average bare
resonance in the system. In Sec.~\ref{sec:WedOct14101637GMT2009} we
will see that when the qubits are detuned, this resonance happens at
$(\omega_1+\omega_2)/2=-\Delta/2$.

Again, although one may expect a triplet in MC, only distorted
singlets are observed in the best of cases due to the broadening and
dispersive parts. In Fig.~\ref{fig:SatJan24165822GMT2009}(c) and (d)
we can see the transition from SSC into MC, at the second vertical
line.

Note that, in this system, the pumping mechanism is equivalent to an
upward decay, due to the ultimate saturation of the qubit and the
symmetry in the schema of levels that they form. The master equation
is symmetrical under exchange of the pump and the decay
($\gamma_i\leftrightarrow P_i$) when the two-levels of both qubit are
inverted ($\ket{0,0}\leftrightarrow \ket{1,1}$ and
$\ket{1,0}\leftrightarrow \ket{0,1}$)~\footnote{That is, pictorially,
  when the structure of levels and transitions are rotated by
  180$^\circ$ or, mathematically, when the rising and lowering
  operators are inverted.}. Consequently, the parameters $z_1$, $z_2$
and $R$, and also the populations of all the levels, are symmetric in
the same way, as it happens with just one qubit. In other systems,
like the linear model, the Jayne-Cummings model or simply a single
harmonic oscillator, the effect of the pump extends upwards to an
infinite number of manifolds while the decay cannot bring the system
lower than the ground state. There is no natural truncation for the
pump (that ultimately leads to a divergence), as there is for the
decay. But with coupled qubit, state $\ket{1,1}$ is the upper
counterpart of $\ket{0,0}$, undergoing a saturation. This implies, for
instance, that the (transient) dynamics in the limit of vanishing
decay is exactly the same as that of vanishing pump and that in such
case we can also apply the manifold method to obtain the right
positions and broadenings as a function of pump, in the same way that
we did as a function of decay only. We only have to take into account
the mentioned symmetry consistently. As long as the dynamics moves
upwards or downwards only, even when intermediate states are coupled,
the manifold picture is suitable. The manifold diagonalization breaks,
however, in the presence of both non-negligible pump and decay. This
is discussed in appendix~\ref{app:SatJan9154344GMT2010}.

\section{Particular cases}
\label{sec:SunFeb1004214CET2009}

\begin{figure}[t]

\centering
\includegraphics[width=0.47\linewidth]{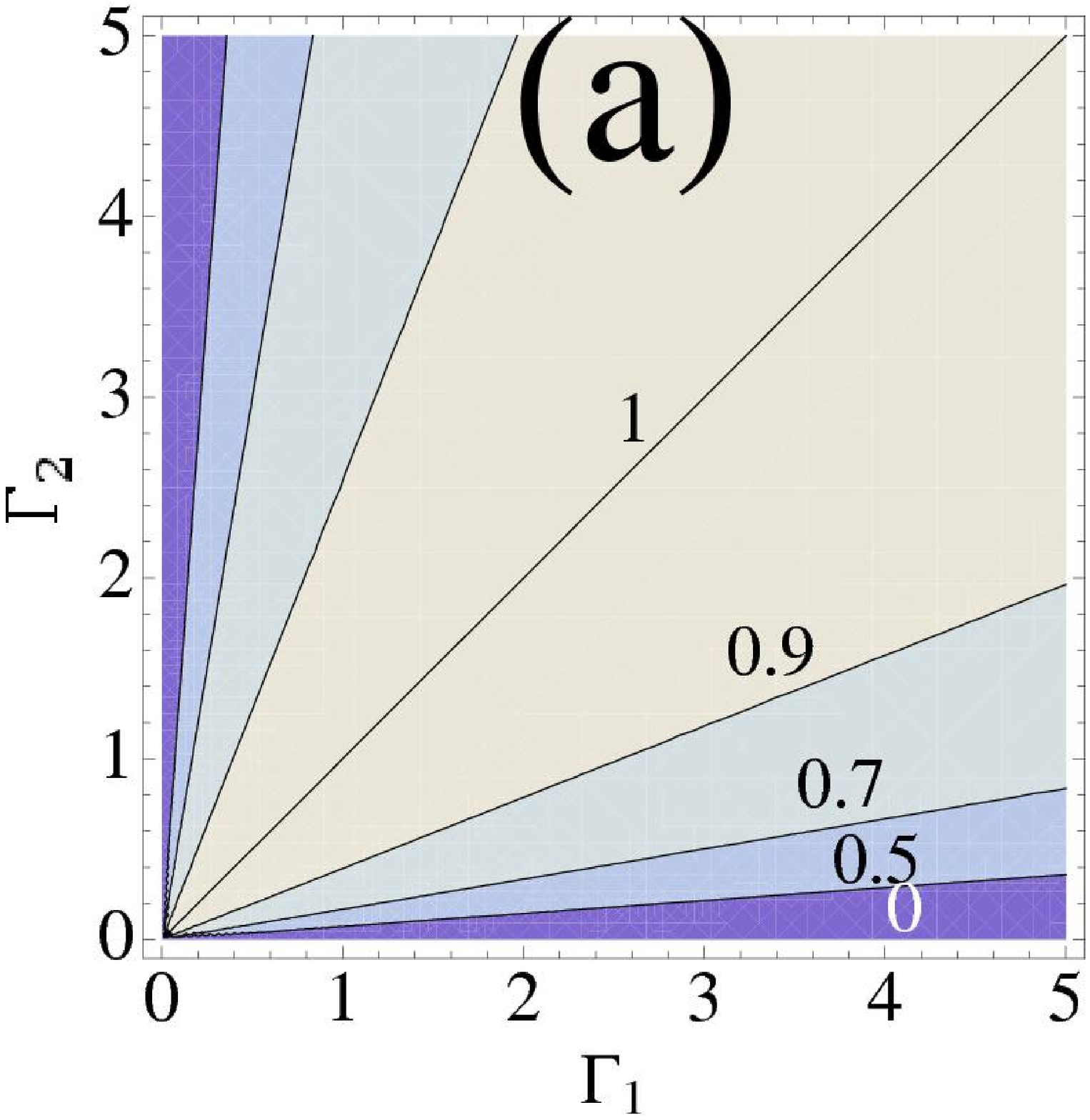}
\includegraphics[width=0.48\linewidth]{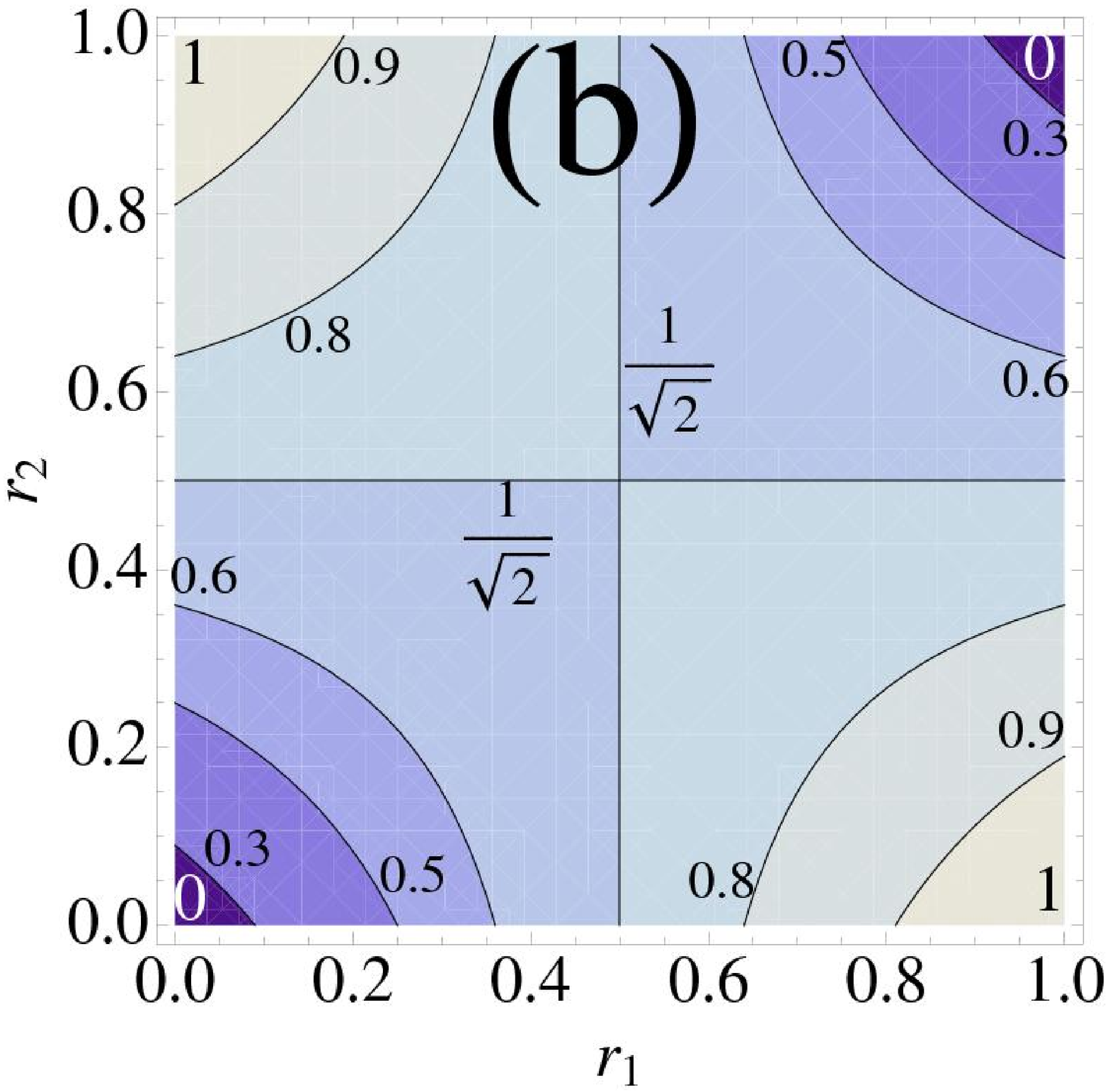}
\caption{Factors contributing to $D^\mathrm{s}$: (a) $\sqrt{\Gamma_1
    \Gamma_2}/[(\Gamma_1+\Gamma_2)/2]$ as a function of
  $\Gamma_1,\Gamma_2$ and (b) $\sqrt{r_1+r_2-2r_1 r_2}$ as a function
  of $r_1,r_2$. The values corresponding to the contour lines are
  marked on the plots. Both functions take values from 0 (dark blue)
  to 1 (light).}
\label{fig:WedOct14122001GMT2009}
\end{figure}

In this section, we illustrate the rather abstract previous
discussions with examples. The symmetry in the decay and pumping rates
determines the effective coupling, emission properties and dressed
states. Let us start by expressing $D^\mathrm{s}$, the magnitude
quantifying such symmetry, in terms of the reservoir parameters
($\Gamma_i$ and $r_i$),
\begin{equation}
  \label{eq:WedOct14100247GMT2009}
  D^\mathrm{s}=\sqrt{2}\frac{\sqrt{\Gamma_1 \Gamma_2}}{(\Gamma_1+\Gamma_2)/2}\sqrt{r_1+r_2-2r_1 r_2}\,.
\end{equation}
In this form, its physical meaning is more clear. There are two
separate factors to discuss: the symmetry in the strength of the
couplings to the reservoirs, given by $\sqrt{\Gamma_1
  \Gamma_2}/[(\Gamma_1+\Gamma_2)/2]$ and plotted in
Fig.~\ref{fig:WedOct14122001GMT2009}(a), and the symmetry in the
nature of the reservoirs given by $\sqrt{r_1+r_2-2r_1 r_2}$,
plotted in Fig.~\ref{fig:WedOct14122001GMT2009}(b).

The coupling $g$ is enhanced when
\begin{equation}
  \label{eq:WedOct14153427GMT2009}
  g< G \leq\sqrt{2}g,\quad\text{that is,}\quad 1<D^\mathrm{s}\leq \sqrt{2}\,,
\end{equation}
which happens when $r_1>1/2$ and $r_2<1/2$ (or the other way
around). This corresponds to the two squared regions with lighter
colors in Fig.~\ref{fig:WedOct14122001GMT2009}(b). Then, the two
reservoirs are of opposite natures: the reservoir of the first qubit
provides excitations ($P_1>\gamma_1$) while the other absorbs them
($P_2<\gamma_2$). If this is accompanied by similar interaction
strengths, $\Gamma_1\sim\Gamma_2$, enhancement occurs. We refer to
these situations as \emph{optimally pumped} and study them in
Sec.~\ref{sec:WedOct14101637GMT2009}.

On the other hand, if the reservoirs are of the similar natures, both
$r_1,r_2\geq 1/2$ or $\leq 1/2$, both providing or absorbing
particles, then the system is \emph{detrimentally pumped}:
\begin{equation}
  \label{eq:WedOct14153709GMT2009}
  0\leq G \leq g,\quad\text{that is,}\quad 0\leq D^\mathrm{s}\leq 1\,.
\end{equation}
It corresponds to the two squared regions with darker colors in
Fig.~\ref{fig:WedOct14122001GMT2009}(b). We study this in
Sec.~\ref{sec:WedOct14131208GMT2009}.

\subsection{Optimally pumped cases: $g<G\leq\sqrt{2}g$}
\label{sec:WedOct14101637GMT2009}

Let us explore the \emph{optimally pumped cases} by considering
parameters on the diagonal $\Gamma_1=\Gamma_2$ in
Fig.~\ref{fig:WedOct14122001GMT2009}(a) together with the antidiagonal
$r_1+r_2=1$ in Fig.~\ref{fig:WedOct14122001GMT2009}(b). The
reservoirs have opposite nature but interact with equal strength with
the qubits. This corresponds to the situation where the decay and
pumping parameters are equal in a crossed way:
\begin{equation}
  \label{eq:ThuMay7104955GMT2009}
  P_1=\gamma_2\,, \quad \text{and} \quad P_2=\gamma_1\,.
\end{equation}
The system has a total input that is equal to the total output,
$P_\mathrm{TOT}=P_1+P_2=\gamma_\mathrm{TOT}=\gamma_1+\gamma_2$, and
also equal Purcell rates, $Q_1=Q_2$. The excited and ground states are
formally equivalent in the dynamics.

\begin{figure}[t]
\centering
\includegraphics[width=0.9\linewidth]{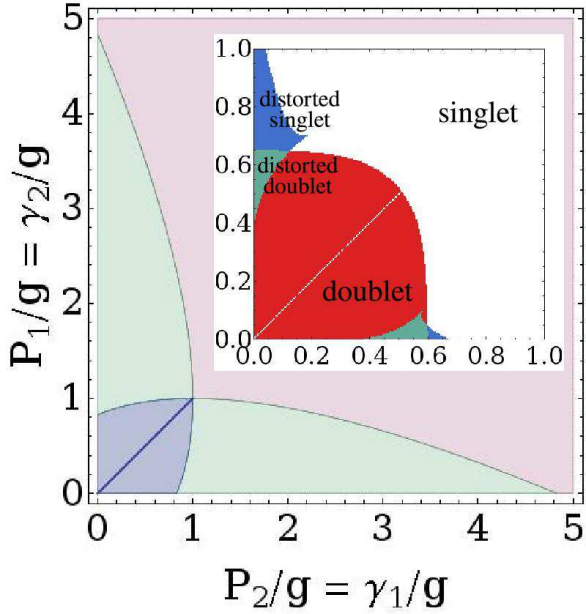}
\caption{(Color online) Phase space of FSC/SSC/MC/WC as function of
  $P_1/g=\gamma_2/g$ and $P_2/g=\gamma_1/g$. The color code is that of
  Fig.~\ref{fig:SatJan24165755GMT2009}. In inset, the possible
  lineshapes of $S_1(\omega)$: a doublet (red), a distorted doublet
  (green), a distorted singlet (blue) and a singlet (white), as
  explained in Table~\ref{tab:SunFeb1041903CET2009}.}
\label{fig:SunFeb1065321CET2009}
\end{figure}

Figure~\ref{fig:SunFeb1065321CET2009} shows the different coupling
regimes accessible with this configuration, as a function of $P_1$ and
$P_2$ with the same color code than in
Fig.~\ref{fig:SatJan24165755GMT2009}.  This configuration is in FSC
only when all parameters are equal, $D^\mathrm{s}=1$, (blue line) and
there is \emph{total symmetry} in the system. Otherwise, one of the
new type of coupling (SSC in blue or MC in green) is realized as the
coupling is effectively improved, $G>g$.  In the inset the type of
spectral shapes that results is shown.

\begin{table}[hbpt]
  \centering
  \begin{tabular}{| c | c | c |c|}
    \hline
    Lineshape & $L_\mathrm{sl}$ & $L_\mathrm{con}$\\
    \hline
    \hline
    singlet&1&2\\
    distorted singlet&1&6\\
    doublet&3&4\\
    distorted doublet&3&8\\
    triplet&5&6\\
    quadruplet&7&8\\
    \hline
  \end{tabular}
  \caption{The lineshapes~$S_1(\omega)$ are defined by two quantities: $L_\mathrm{sl}$ is the number of times that~$S_1(\omega)$ changes slope, that is, the number of real solutions to the equation~$dS_1(\omega)/d\omega=0$; $L_\mathrm{con}$ is the number of times that~$S_1(\omega)$ changes concavity, that is, the number of real solutions to the equation~$d^2S_1(\omega)/d\omega^2=0$.}
\label{tab:SunFeb1041903CET2009}
\end{table}

\begin{figure}[t]
\centering
\includegraphics[width=\linewidth]{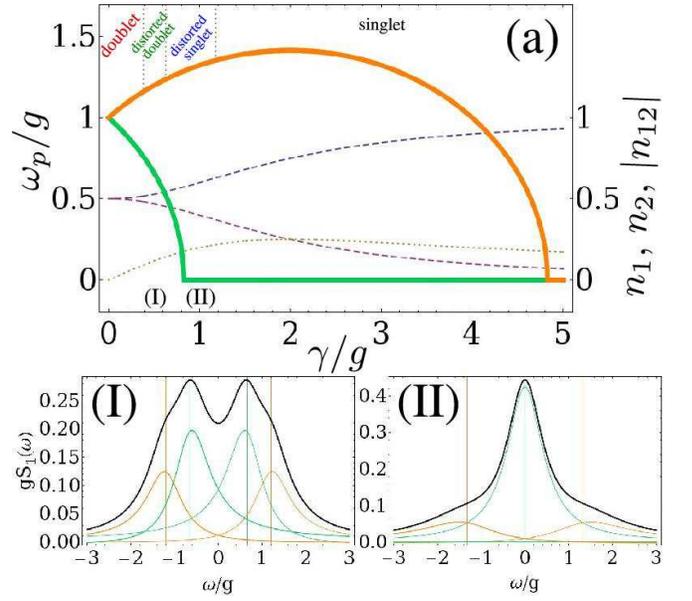}
\caption{(Color online) (a) Positions of the peaks ($\omega_A$ in
  thick green, $\omega_B$ in thick orange) and populations ($n_1$ in
  dashed blue, $n_2$ in dashed purple, $|n_\mathrm{corr}|$ in dotted
  brown) as function of $P_1/g=\gamma_2/g=\gamma/g$, for
  $P_2=\gamma_1=0$. The system goes from FSC (at 0) to SSC, to MC, to
  WC, while the lineshape of the spectra changes as coded in
  Table~\ref{tab:SunFeb1041903CET2009}. The most interesting
  lineshapes, that can only appear in SSC and MC, are the distorted
  doublet (I) and singlet (II). The total spectra (in black) is
  decomposed in inner (green) and outer (orange) peaks coming from the
  transitions sketched in Fig.~\ref{fig:FriJan23211556GMT2009}(d).}
\label{fig:SunFeb1041856CET2009}
\end{figure}

The vertical axis in Fig.~\ref{fig:SunFeb1065321CET2009}, with
$P_1=\gamma_2=\gamma$ and $P_2=\gamma_1=0$, is illustrative of all the
possible coupling regions and lineshapes. This is the extreme
situation of optimal pumping where the two reservoirs interact equally
strongly with the two qubits ($\Gamma_1=\Gamma_2=\gamma$) and have
completely opposite natures: one only providing particles ($r_1=1$)
and the other only absorbing them ($r_2=0$). At this point, there is
maximum renormalization of the coupling, $G=\sqrt{2}g$
($D^\mathrm{s}=\sqrt{2}$) as both factors in
Eq.~(\ref{eq:WedOct14100247GMT2009}) are maximum. The populations and
mean values read:
\begin{subequations}
  \label{eq:SunFeb1093915CET2009}
  \begin{align}
    &n_2=\frac{2}{4+(\gamma/g)^2}\,,\quad n_1=1-n_2\,,\\
    &n_{11}=\frac{1}{4+(\gamma/g)^2}\neq n_1 n_2\,,\\
    &n_\mathrm{corr}=-i\frac{\gamma/g}{4+(\gamma/g)^2}\,.
  \end{align}
\end{subequations}
The two qubits share one excitation only. The Rabi frequency also
simplifies to $R=ig$ (as $\Gamma_-=0$), and
\begin{equation}
  \label{eq:ThuOct15155615GMT2009}
  z_{1,2}=\sqrt{g^2-(\gamma/2)^2\mp g\gamma}\,.
\end{equation}
In Fig.~\ref{fig:SunFeb1041856CET2009}, we can see some of these
magnitudes varying in the different regimes as a function of
$\gamma/g$. In the linear regime limit, $\gamma\ll g$, there is FSC
with all the levels equally populated ($n_1=n_2=1/2$, $n_{11}=1/4$)
and $n_\mathrm{corr}=-i(\gamma/g)/4$. As soon as there is pumping, the
SSC opens a ``bubble'' in the eigenenergies with the splitting of
inner and outer peaks. The transition into MC, with the collapse of
the inner peaks, takes place at $\gamma=2(\sqrt{2}-1)g$, and the
transition into WC, closing the bubble, takes place at
$\gamma=2(\sqrt{2}+1)g$. The maximum of $z_2=\sqrt{2}g$ (in orange)
takes place at $\gamma=2g$, when the coherence $|n_\mathrm{corr}|=1/4$
is maximum. This is a special point where the splitting of the dressed
modes is the largest possible, $2\sqrt{2}g$ (even though the lineshape
remains a singlet). Finally, when the coupling becomes very weak,
$\gamma\gg g$, the first dot saturates, quenching the exchange of
excitation between the qubit and $n_2=n_{11}=n_\mathrm{corr}=0$.
 
Although the underlying physics of coupling is very rich and complex,
the spectra do not acquire distinctively marked lineshapes (such as
well resolved triplets of quadruplets): a doublet in SSC only gets
distorted, but this is unambiguously due to the underlying quadruplet
structure, as shown in Fig.~\ref{fig:SunFeb1041856CET2009}(I), and
also a singlet gets distorted due to the underlying triplet structure,
in (II). Before reaching WC, the spectrum has become a plain
singlet. The way to distinguish mathematically the different possible
shapes and their origin in underlying triplets and quadruplets is by
counting zeros of first and second derivatives, as given in
Table~\ref{tab:SunFeb1041903CET2009}, of which only the four first
lines are realized in the coupled qubit.

\begin{figure}[t]
\centering
\includegraphics[width=\linewidth]{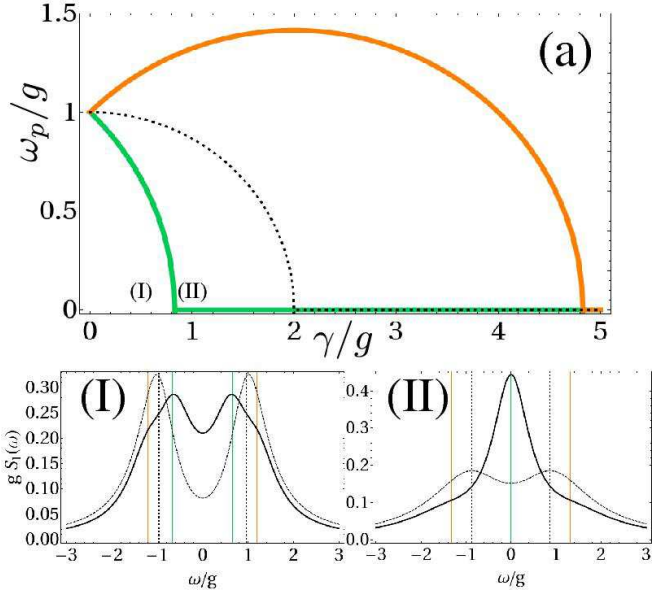}
\caption{(Color online) Comparison between the results of
  Fig.~\ref{fig:SunFeb1041856CET2009} (solid lines) and those obtained
  for a 4LS (dashed black lines). The positions of the peaks composing
  the spectra, are marked with vertical lines in (I) and (II). In a
  4LS, only plain doublets and singlets arise. The populations are the
  same for both cases. Different types of splitting lead to dramatic
  differences in the spectral shapes.}
\label{fig:MoNov23201007WET2009}
\end{figure}

In Fig.~\ref{fig:MoNov23201007WET2009} we make the comparison between
the results discussed in Fig.~\ref{fig:SunFeb1041856CET2009} and those
obtained for a 4LS (without cross Lindblad terms), where the system
undergoes a standard transition SC-WC as $\gamma$ increases. From
Sec.~\ref{sec:MoNov23201725WET2009}, we know that, although the
populations do not change from those in
Eqs.~(\ref{eq:SunFeb1093915CET2009}), the position of the four peaks
differ, as well as the coupling regimes. In this case the parameters
$z$ converge to:
\begin{equation}
  \label{eq:MoNov23223002WET2009}
  z_{1,2}\rightarrow \sqrt{g^2-(\gamma/2)^2}\,.
\end{equation}
This corresponds to a splitting intermediate between those of the
inner and outer peaks, as we can see in dashed black in
Fig.~\ref{fig:MoNov23201007WET2009}(a). Although the 4LS splitting is
smaller than the splitting between outer peaks (in orange), the SC
doublet in the spectrum is better resolved than in the case of two
qubit, due to the large intensity of the inner peaks.
\begin{figure}[tb]
\centering
\includegraphics[width=\linewidth]{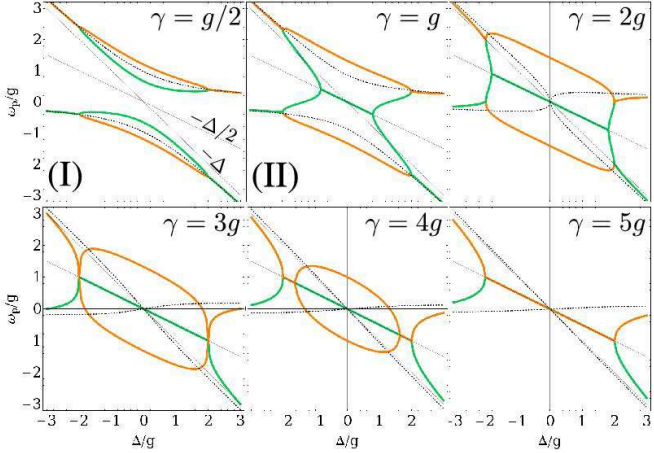}
\caption{(Color online) (a) Positions of the peaks ($\omega_{A,C}$ in
  thick green, $\omega_{B,D}$ in thick orange) as a function of
  detuning for six cases in Fig.~\ref{fig:SunFeb1041856CET2009} with
  the specified parameter $\gamma$. The positions in the absence of
  crossed terms appear in dashed black and in the absence of coupling
  in thin black, for comparison. The first two plots are those labeled
  (I) and (II) in previous Figures.}
\label{fig:ThuAug6190254GMT2009}
\end{figure}

\begin{figure}[t]
\centering
\includegraphics[width=\linewidth]{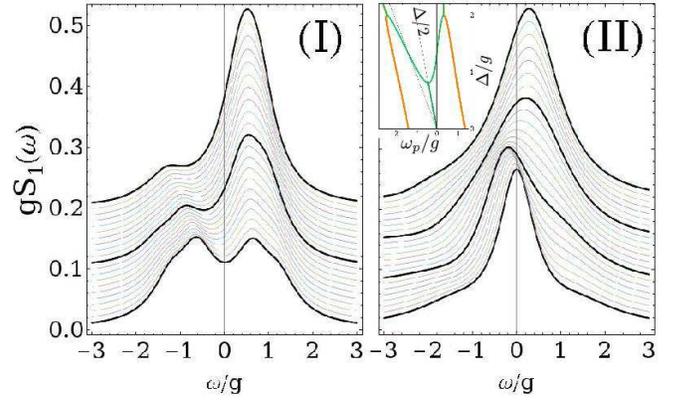}
\caption{(Color online) Spectra for varying positive detuning
  (anticrossings) in SSC (I) and in MC (II) for the corresponding
  cases in Fig.~\ref{fig:MoNov23201007WET2009}. The inset in (II)
  shows the positions of the three/four peaks composing the spectra as
  a function of detuning. Close to resonance, the inner peaks converge
  to the average transition energy $\omega_{11}/2=\Delta/2$. At large
  detunings, the first dot (that is pumped) emits at $\omega_1=0$ and
  dominates over the second dot (that decays) with emission at
  $\omega_2=-\Delta$.}
\label{fig:FriFeb13174253CET2009}
\end{figure}

Fig.~\ref{fig:ThuAug6190254GMT2009} displays the behaviors of all the
resonances as a function of detuning for six cases in
Fig.~\ref{fig:SunFeb1041856CET2009} (see insets), including (I)
$\gamma=g/2$ and (II) $\gamma=g$. Inner and outer peaks superimpose
whenever there are only one or two resonances. Otherwise, green lines
represent inner resonances and orange lines the outer ones. Dashed
black lines represent the case of a 4LS, which simply splits in two
lines close to resonance in SC (the transition to WC happens at
$\gamma=2g$). In thin black, the bare energies ($\omega_1=0$,
$\omega_2=-\Delta$) are plotted as a reference, and recovered in all
cases very far from resonance or well into WC. The 4LS resonances
converge to the bare energies at all detunings almost as soon as they
enter WC. However, the resonances of two qubits, when closing in WC
(MC in the case of inner resonances), collapse to the average between
the bare energies $(\omega_1+\omega_2)/2=-\Delta/2$ instead. At
resonance these two behaviors are equivalent, but at small detunings
they are clearly very different. We will discuss the reason for this
puzzling saturation at $-\Delta/2$ instead of $-\Delta$ in
Sec.~\ref{sec:WedJul29173348GMT2009}. Eventually, at very large
$\gamma$, the bare energies will be recovered both by inner and outer
peaks.

The anticrossing that the lineshapes form when detuning between the
modes is varied from zero to some detuning~$\Delta_\mathrm{max}$, is
also peculiar.  In Fig.~\ref{fig:FriFeb13174253CET2009}(I) and (II) we
can see that the distorted doublet and singlet keep their features up
to $\Delta_\mathrm{max}=g$ and $\Delta_\mathrm{max}=2g$ (resp.). The
resonances in Fig.~\ref{fig:FriFeb13174253CET2009} are not equally
present in the emission. In both cases, at large detuning, the
emission at $\omega_1=0$ of the first qubit is dominant over that of
the second qubit at $\omega_2=-\Delta$ because the first qubit is
being pumped and the second only dissipates the excitation. Note that,
the singlet in (II) slightly oscillates to the left before joining the
origin at large detuning. This is a clear signature that the central
peak is not simply a feature of WC, but of a more complicated MC
eigenstate structure, as can be seen in the inset. This leftwards
shift is a consequence of the splittings saturating at $-\Delta/2$
instead of the bare energies ($0$ and $-\Delta$). This shift is even
more evident when both inner and outer peaks saturate at some
detuning, like in case $\gamma=4g$.

\subsection{Detrimentally pumped cases: $0<G\leq g$}
\label{sec:WedOct14131208GMT2009}

To explore the \emph{detrimentally pumped} cases, we consider
parameters on the diagonal $r_1=r_2=r$ in
Fig.~\ref{fig:WedOct14122001GMT2009}(b), where the reservoirs have the
same nature.

If the reservoirs also have the same interaction strength with the
qubit, $\Gamma_1=\Gamma_2$, the two qubits become indistinguishable. This
corresponds to equal dissipation and pump,
\begin{equation}
  \label{eq:ThuMay7114051GMT2009}
  \gamma_1=\gamma_2=\gamma \quad \text{and} \quad P_1=P_2=P\,.
\end{equation}
The symmetry in the system is not total as the income-outcome flows of
excitation may not be balanced, $\gamma_\mathrm{TOT}\neq
P_\mathrm{TOT}$. The steady state populations are the same as for the
uncoupled case with $n_1=n_2=P/(\gamma+P)$, $n_{11}= n_1 n_2$,
$n_\mathrm{corr}=0$, as happens with any two indistinguishable coupled
modes (like, for instance, harmonic oscillators). This does not mean
that the qubit are uncoupled, in fact, the system can be considered
always in strong coupling (FSC) with
\begin{subequations}
  \label{eq:SunFeb1111540CET2009}
  \begin{align}
    &R=\frac{|\gamma-P|}{\gamma+P}g=2|r-\frac{1}{2}|g\,, \\
    &D^\mathrm{s}=\frac{\sqrt{\gamma
        P}}{(\gamma+P)/2}=2\sqrt{r(1-r)}\,.
  \end{align}
\end{subequations}
If $P\gg\gamma$ ($r=1$), or the other way around ($r=0$), the symmetry
between the flows (inwards and outwards) is completely broken and
$D^\mathrm{s}=0$, although, at the same time, the splitting in dressed
modes is maximum~$R=g$ (although not enhanced). 

On the other hand, if $P=\gamma$ ($r=1/2$), the symmetry is total
($D^\mathrm{s}=1$). Here, all the levels are equally populated
($n_1=n_2=1/2$, $n_{11}=1/4$) and the Rabi frequency vanishes
$R=0$. The SC/WC regimes become conventional with
$z_{1,2}=\sqrt{g^2-\gamma^2}$. The SC condition is simply given by
$g>\gamma$. This particular situation is depicted in
Fig.~\ref{fig:SunFeb1065321CET2009} with a thick dark blue line in SC
that goes into WC when the parameters equal~1.

A second possibility that results in degrading the effective coupling,
still considering reservoirs of the same nature ($r_1=r_2=r$), is when
$r=1/2$, which means:
\begin{equation}
  \label{eq:ThuMay7114911GMT2009}
  \gamma_1=P_1=\Gamma_1/2 \quad \text{and} \quad \gamma_2=P_2=\Gamma_2/2\,.
\end{equation}
Although the parameters may be different, $\Gamma_1\neq\Gamma_2$,
there is compensation between the flows (inwards and outwards)
$\gamma_\mathrm{TOT} = P_\mathrm{TOT}$. This configuration arises when
both qubits are in a thermal bath of infinite temperature. The
populations are also those of the uncoupled system, $n_1=n_2=1/2$,
$n_{11}= n_1 n_2$, $n_\mathrm{corr}=0$. The Rabi and degree of
symmetry read
\begin{equation}
  \label{eq:SunFeb1114521CET2009}
  R=\frac{|\Gamma_-|}{\Gamma_+}\sqrt{g^2-\Gamma_+^2} \quad \text{and} \quad D^\mathrm{s}=\frac{\sqrt{\Gamma_1\Gamma_2}}{(\Gamma_1+\Gamma_2)/2} \,.
\end{equation}
The FSC condition reads now~$g>\Gamma_+$.

\section{Dressed states}
\label{sec:WedJul29173348GMT2009}

Now that we have described the new regimes appearing in this system
due to the incoherent pumping, SSC and MC, and their spectral
features, we can discuss their origin.

We have pointed out that the manifold picture, based on the
Hamiltonian dynamics, breaks in presence of non-negligible pumping. To
define dressed states in such an essentially non-Hamiltonian system,
one must recourse to another formalism, presented in this section,
based on studying the flow of coherence in the system. This allows a
reconstruction of dressed states that recovers the manifold picture
for vanishing pump and extends it otherwise.

To analyze how coherence flows in the system, let us address the
dynamics of coherence between states, given by off-diagonal elements
of the density matrix~$\rho$. We thus consider the vector of the
relevant transitions:
\begin{equation}
  \label{eq:MonNov16165000GMT2009}
  \mathbf{v}_\mathrm{coh}=
    \begin{pmatrix}
\mean{u_1}\\
\mean{l_1}\\
\mean{u_2}\\
\mean{l_2}
    \end{pmatrix}=
    \begin{pmatrix}
\bra{1,1}\rho\ket{0,1}\\
\bra{1,0}\rho\ket{0,0}\\
\bra{1,1}\rho\ket{1,0}\\
\bra{0,1}\rho\ket{0,0}
    \end{pmatrix}\,.
\end{equation}
The dynamics of this vector is ruled by a matrix
$\mathbf{M}_\mathrm{coh}$:
\begin{equation}
  \label{eq:SatJan9144935GMT2010}
  d\mathbf{v}_\mathrm{coh}/dt
  =-\mathbf{M}_\mathrm{coh}\mathbf{v}_\mathrm{coh}
\end{equation}
that reads:
\begin{equation}
  \label{eq:SunNov15201047GMT2009}
  \mathbf{M}_\mathrm{coh}=
    \begin{pmatrix}
      \frac{\Gamma_1}{2}+\gamma_2 & -P_2 & -ig & 0\\
      -\gamma_2 & \frac{\Gamma_1}{2}+P_2  & 0 & ig \\
      -ig & 0 & \frac{\Gamma_2}{2}+\gamma_1-i\Delta & -P_1\\
      0 & ig & -\gamma_1 & \frac{\Gamma_2}{2}+P_1-i\Delta
    \end{pmatrix}\,.
\end{equation}
The diagonal terms in $\mathbf{M}_\mathrm{coh}$ give the decay rate of
each coherence and the off-diagonal terms, the rates at which
coherences is transferred from one transition to another. Its
eigenvalues recover the positions and broadenings of the spectrum
peaks in Eq.~(\ref{eq:TueDec23152835CET2008}), with eigenstates that I
note $\mean{\mathrm{T}_p}$, $p=A,B,C,D$. Therefore, the single time
dynamics of $\mean{\mathrm{T}_p}$ is free:
\begin{equation}
  \label{eq:MoDec7195257WET2009}
 \mean{\mathrm{T}_p(t)}=e^{-(i\omega_p+\gamma_p/2)(t-t_0)}\mean{\mathrm{T}_p(t_0)}\,,
\end{equation}
and their spectral shape is a Lorentzian:
\begin{equation}
  \label{eq:WedNov25133917WET2009}
  s_{T_p}(\omega)=\mean{\ud{\mathrm{T}}_p(\omega)\mathrm{T}_p(\omega)}=\frac{\mean{\ud{\mathrm{T}}_p\mathrm{T}_p}}{\pi}\frac{\gamma_p/2}{(\gamma_p/2)^2+(\omega-\omega_p)^2}\,.
\end{equation}

In this way, one can reconstruct the dressed states from the above
expressions by inspection of the $\langle T_p\rangle$.

First, let us consider the simplest case of Hamiltonian dynamics
(without pumping and decay), for which we already obtained the dressed
states through the manifold picture (by diagonalization of the
Hamiltonian with only direct coupling $g$). The following discussion
will thus reconstruct Fig.~\ref{fig:FriJan23211556GMT2009}(b). The
eigenenergy $\omega_A=\omega_B=+ig$ corresponds to the two
eigenstates:
\begin{subequations}
  \label{eq:SunJan10134247GMT2010}
  \begin{align}
    \label{eq:SunJan10134126GMT2010}
    &\mean{\mathrm{T}_A}=\mean{d_1+d_2}\quad\text{(lower transition $\ket{0,0}\leftrightarrow \ket{+}$)}\,,\\
    &\mean{\mathrm{T}_B}=\mean{u_2-u_1}\quad\text{(upper transition
      $\ket{-}\leftrightarrow \ket{1,1}$)}\,.
  \end{align}
\end{subequations}
The eigenenergy $\omega_C=\omega_D=-ig$ corresponds to the counterpart
eigenstates:
\begin{subequations}
  \label{eq:SunJan10134431GMT2010}
  \begin{align}
    \label{eq:SunJan10134327GMT2010}
    &\mean{\mathrm{T}_C}=\mean{d_1-d_2}\quad\text{(transition $\ket{0,0}\leftrightarrow \ket{-}$)}\,,\\
    &\mean{\mathrm{T}_D}=\mean{u_2+u_1}\quad\text{(transition
      $\ket{+}\leftrightarrow \ket{1,1}$)}\,.
  \end{align}
\end{subequations}
Therefore, the system resonances $\langle T_p\rangle$ give away the
existence of the dressed states:
\begin{equation}
  \label{eq:SunJan10134700GMT2010}
  \ket{\pm}\propto\pm \ket{1,0}+\ket{0,1}
\end{equation}
(left not normalized for convenience here and further-on). We can
clearly see by inspection of $\mathbf{M}_\mathrm{coh}$ how the direct
coupling transfers coherence between the two lower transitions $l_i$,
on the one hand, and the two upper transitions $u_i$, on the other
hand. In this case, coherence can follow a closed path forming loops
that flow always at the same rate $2g$ (in between $u_1\leftrightarrow
u_2$ and $d_1\leftrightarrow d_2$) and give rise to Rabi oscillations
and dressed states.

In the case of a 4LS, the situation is not much different, even when
adding pump and decay: dressed states are also identified through the
same procedure, diagonalizing the corresponding matrix without
incoherent crossed terms:
\begin{equation}
  \label{eq:FriNov20170739WET2009} 
  \mathbf{M}_\mathrm{coh}=
    \begin{pmatrix}
      \frac{\Gamma_1}{2}+\gamma_2 & 0 & -ig & 0\\
      0 & \frac{\Gamma_1}{2}+P_2  & 0 & ig \\
      -ig & 0 & \frac{\Gamma_2}{2}+\gamma_1-i\Delta & 0\\
      0 & ig & 0 & \frac{\Gamma_2}{2}+P_1-i\Delta
    \end{pmatrix}\,.
\end{equation}
In this case, the eigenenergies we found in
Eq.~(\ref{eq:WedAug5140534GMT2009}) correspond to the same combination
of transitions,
Eqs.~(\ref{eq:SunJan10134247GMT2010}-\ref{eq:SunJan10134431GMT2010})
with:
\begin{equation}
  \label{eq:FriNov20174227WET2009}
  \ket{\pm}\propto\frac{\tilde\Gamma_-\pm i\tilde R}{g} \ket{1,0}+i\ket{0,1}\,.
\end{equation}
If $\tilde\Gamma_-=0$ ($\tilde R=g$), we recover the dressed states
without decoherence. Up to the transition into WC at
$|\tilde\Gamma_-|=g$, both coefficients in
Eq.~(\ref{eq:FriNov20174227WET2009}) are complex numbers with unit
norm and, therefore, they give rise to a balanced contribution of the
two states $\ket{1,0}$ and $\ket{0,1}$ equal to $1/2$. Their mean
energy, $\bra{\pm}H\ket{\pm}=\pm \tilde R$, is exactly the one
appearing in the spectral resonances $\omega_p$. The straightforward
identification of dressed states, in the same way as when there is
only direct coupling, makes it possible to describe the dynamics in
the manifold picture as explained in
appendix~\ref{app:SatJan9154344GMT2010}.

\begin{figure}[t]
\centering
\includegraphics[width=0.8\linewidth]{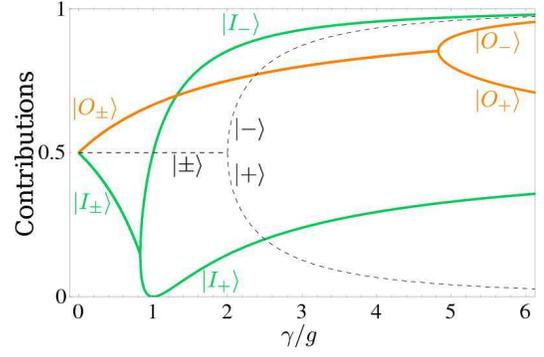}
\caption{(Color online) (a) Contributions from state $\ket{1,0}$ to
  the dressed states appearing in an optimally pumped case:
  $P_1=\gamma_2=\gamma$, $P_2=\gamma_1=0$. The $x$-axis is $\gamma/g$,
  as in previous Figs.~\ref{fig:SunFeb1041856CET2009} and
  ~\ref{fig:MoNov23201007WET2009} describing the same case. In thin
  dashed black lines, the contribution to the 4LS dressed states
  $\ket{\pm}$. In thick lines, the qubit dressed states: orange for
  $\ket{O_\pm}$ and green for $\ket{I_\pm}$. The splitting of each
  pair of lines $\pm$ from equal to different contributions, marks
  their transition into WC. All the ``$-$'' states tend towards
  $\ket{1,0}$ in the very weakly coupled regime.}
\label{fig:MoNov30120211WET2009}
\end{figure}

When the system is in WC ($\tilde R=i|\tilde R|$), the dressed states
become essentially the bare ones, $\ket{1,0}$ and $\ket{0,1}$, since
$|\tilde\Gamma_-|/g\rightarrow \infty$. That is, in this regime, the
populations become less and less balanced, as we can see in the
example of Fig.~\ref{fig:MoNov30120211WET2009}, where the probability
to find the system in $\ket{1,0}$ is plotted in thin dashed for both
$\ket{\pm}$.

On the other hand, in the case of two qubits which are only pumped and
decaying, with no direct coupling, the eigenstates of
$\mathbf{M}_\mathrm{coh}$ are combinations of the transitions
corresponding to each qubit: $\mean{P_2 u_1+\gamma_2 d_1}$, $\mean{P_1
  u_2+\gamma_1 d_2}$, $\mean{u_1-d_1}$ and $\mean{u_2-d_2}$. No new
dressed states arise although coherence is being transferred
(incoherently) between two transitions.

From these three limiting situations, we learn that direct coupling is
essential for the appearance of dressed states but also that pump and
decay can transfer coherence in the directions $u_1\leftrightarrow
d_1$ and $u_2\leftrightarrow d_2$, in an incoherent continuous way. In
spite of this, they do not induce oscillations in the dynamics. Even
in the most symmetrical case where $\gamma_1=P_1$ and $\gamma_2=P_2$,
pump and decay rather move around coherence in an incoherent way,
basically disrupting the Rabi oscillations, as we observed at the end
of Sec.~\ref{sec:WedOct14131208GMT2009}: $G\leq g$.

\begin{figure}[t]
\centering
\includegraphics[width=0.6\linewidth]{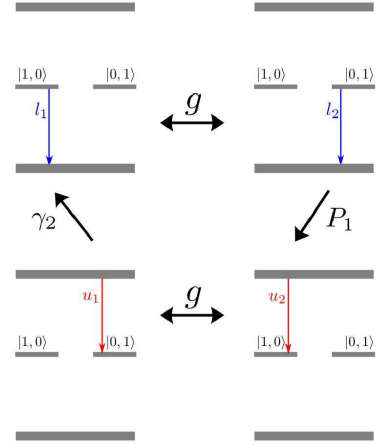}
\caption{(Color online) Circulation of coherence among the transitions
  between bare states for two qubits in the case where
  $\gamma_1=P_2=0$. A loop is formed that involves all the
  transitions, thanks to the pumping and decay terms that participate
  to the flow of coherence in a constructive way. The direct coupling
  goes both in tune with the loop or against it. When
  $\gamma_2=P_1=2g$ the transfer is optimal.}
\label{fig:MoNov30120210WET2009}
\end{figure}

Exceptionally, if the coherence transfer induced by pump and decay
happens in a constructive, rather than independent, way as compared to
the one induced by the direct coupling, Rabi oscillations and dressed
states can be enhanced. This is the case of the optimally pumped
configurations studied in Sec.~\ref{sec:WedOct14101637GMT2009}. Let us
analyze, for instance, the best situation, $\gamma_2=P_1=0$ and
$\gamma_1=P_2=\gamma$, where $G=\sqrt{2}g$. In this case, the
coherence transfer in the system can take place in a loop that
involves not only the direct coupling but also incoherent processes,
as sketched in Fig.~\ref{fig:MoNov30120210WET2009}. This results in
the appearance of four different resonances in the system, that can be
written in terms of two pairs of new intermediate dressed states,
\begin{subequations}
  \label{eq:MoNov30140813WET2009}
  \begin{align}
    &\ket{I_{\pm}}\propto \frac{\gamma/2\pm i z_1}{g}\ket{1,0}+i\ket{0,1}\,,\\
    &\ket{O_{\pm}}\propto \frac{\gamma/2\pm i z_2}{g}\ket{1,0}+i\ket{0,1}\,,
  \end{align}
\end{subequations}
named after \emph{inner} and \emph{outer} eigenenergies. They
correspond to $\gamma \pm i z_1$ and $\gamma \pm i z_2$, respectively,
where $z_{1,2}$ are those in Eq.~(\ref{eq:ThuOct15155615GMT2009}). The
level scheme that these new coherence loops produce is plotted in
Fig.~\ref{fig:FriJan23211556GMT2009}(d). In SSC, the dressed states
appear in the eigentransitions as:
\begin{subequations}
  \label{eq:SatNov21193219WET2009}
  \begin{align}
    &\mean{\mathrm{T}_A}=\mean{\ket{I_-}\bra{1,1}+i\ket{0,0}\bra{I_+}}\,,\\
    &\mean{\mathrm{T}_B}=\mean{\ket{O_-}\bra{1,1}-i\ket{0,0}\bra{O_+}}\,,\\
    &\mean{\mathrm{T}_C}=\mean{\ket{I_+}\bra{1,1}+i\ket{0,0}\bra{I_-}}\,,\\
    &\mean{\mathrm{T}_D}=\mean{\ket{O_+}\bra{1,1}-i\ket{0,0}\bra{O_-}}\,.
  \end{align}
\end{subequations}
The normalization of these dressed states also depends on the coupling
region. In SSC, inner states are normalized by dividing by
$\sqrt{g(2g-\gamma)}$ and outer by $\sqrt{g(2g+\gamma)}$, instead of
$\sqrt{2}g$ as was the case of the 4LS. These states do not have a
balanced contribution from the bare states due to the saturation
brought by pump and decay. The inner states have a smaller
contribution from the first qubit than the outer states:
\begin{subequations}
  \label{eq:SatNov21193219WET20092}
  \begin{align}
    &|\bra{1,0}I_\pm\rangle|^2=\frac{1}{2}\big(1-\frac{\gamma}{2g-\gamma}\big)\,,\\
    &|\bra{1,0}O_\pm\rangle|^2=\frac{1}{2}\big(1+\frac{\gamma}{2g+\gamma}\big)\,.
  \end{align}
\end{subequations}
The 4LS dressed states correspond to an intermediate situation between
the inner and outer dressed states in terms of populations (equal to
$1/2$ for both qubit) and energy (see
Fig.~\ref{fig:ThuAug6190254GMT2009}). This appears clearly in
Fig.~\ref{fig:MoNov30120211WET2009} where the first qubit contribution
to each dressed state is plotted as a function of $\gamma/g$.

The two transitions composing each of the eigentransitions in
Eq.~(\ref{eq:MoNov30140813WET2009}) have the same energy, which does
not coincide with the average Hamiltonian value as soon as decoherence
appears: $\bra{I_\pm}H\ket{I_\pm}=\pm\Re{(z_1)}/[1-\gamma/(2g)]$ and
$\bra{O_\pm}H\ket{O_\pm}=\pm\Re{(z_2)}/[1+\gamma/(2g)]$.

When entering WC, the splittings between each pair of dressed states
close, while the eigentransitions are linked each to a single state:
\begin{subequations}
  \label{eq:WedNov25180213WET2009}
  \begin{align}
    &\mean{\mathrm{T}_A}=\mean{\ket{I_-}\bra{1,1}+i\ket{0,0}\bra{I_-}}\,,\\
    &\mean{\mathrm{T}_B}=\mean{\ket{O_-}\bra{1,1}+i\ket{0,0}\bra{O_-}}\,,\\
    &\mean{\mathrm{T}_C}=\mean{\ket{I_+}\bra{1,1}+i\ket{0,0}\bra{I_+}}\,,\\
    &\mean{\mathrm{T}_D}=\mean{\ket{O_+}\bra{1,1}+i\ket{0,0}\bra{O_+}}\,.
  \end{align}
\end{subequations}
Note that the inner peaks are always the most intense ones and that,
when entering the WC, it is the ``$-$'' states that dominate the
spectrum due to the saturation of the system into their limiting
state $\ket{1,0}$.

\section{Conclusions}
\label{sec:SatMay19212524CEST2007}

I described fully analytically the weak and strong coupling of two
qubits in presence of dissipation and pumping. This configuration is
realized by two two-level systems that commute. This was contrasted
with other models of quantum coupling between two modes, namely, two
bosons (that commute) and two fermions (that anticommute). I also
addressed the case of a four-level system (4LS), which has the same
level structure.

Whereas the coupling of bosons, fermions and a 4LS manifest a
conventional phenomenology, well known from the normal coupling of two
modes and differing only in bosonic or fermionic effective parameters,
I showed that cross-Lindblad terms in the two coupled qubits give rise
to a rich and complex new nomenclature of coupling. Namely, beyond
standard strong coupling (SC) where two dressed states emerge from
quantum superposition of the bare states, I found that four dressed
states, $\ket{I_\pm}$ and $\ket{O_\pm}$, can be realized in the system
with two different Rabi splittings, as a result of pumping and decay
establishing new paths of coherence flow in the system. This is a new
paradigm of strong coupling that leads to new regimes that I called
``Second order Strong Coupling'' (SSC), when the four dressed states
are split (in two pairs), and ``Mixed Coupling'' (MC), when one pair
of dressed states, the inner $\ket{I_\pm}$, have closed. In this wider
picture, I called ``First order Strong Coupling'' (FSC) the
conventional type of strong coupling with only one pair of dressed
states. These results show and spell out the considerable complexity
of coherence transfer in other systems such as the Jaynes-Cumming
model in presence of incoherent pumping, that are not amenable to
analytical treatment.

\begin{acknowledgments}
  I thank F.~P.~Laussy for fruitful discussions and critical reading
  of the manuscript and T.~Ostatnick\'y for useful comments. This
  research was supported by the Newton Fellowship program.
\end{acknowledgments}


\appendix

\section{Interactions and dephasing}
\label{app:TueDec29173303GMT2009}

If the two qubits are close enough in space, they may interact. The
energy levels are affected by the interaction (Coulomb interaction in
the case of atoms or excitons) and the Hamiltonian must be corrected
with the term:
\begin{equation}
  \label{eq:SunJul5225443GMT2009}
  H_\mathrm{b}=-\chi \ud{\sigma_1}\sigma_1\ud{\sigma_2}\sigma_2\,,
\end{equation}
where~$\chi$ is the binding energy between two excitations ($\chi>0$
corresponds to attraction). If $\chi \ll \omega_{1,2}$, interactions
have a negligible effect on the populations and statistics, but not on
the spectral features, starting with
Eq.~(\ref{eq:FriJan8100155GMT2010}) becoming
$\omega_{11}=\omega_1+\omega_2-\chi$. The dynamics in the presence of
interactions is derived in appendix~\ref{app:MonJul6000455GMT2009}
although the expressions for spectrum and dressed states are not
presented in the text as they are not analytical in the most general
case, and bring little to the discussion. The main effect that it
produces is that upper and lower transitions can be identified thanks
to the different energy of the upper transitions and this was used to
check the consistency of the results. In FSC and WC, the difference is
already present due to the broadenings associated to each transition
and interactions only accentuate it. However, in SSC and MC,
interactions break the symmetry between upper and lower transitions
(e.g., $\omega_A$ was linked to both upper or lower transition from
inner states in Fig.~\ref{fig:FriJan23211556GMT2009}(d)) depending on
the sign of~$\chi$:
\begin{itemize}
\item if $\chi>0$, $\omega_{A,D}$ correspond to the upper transitions
  and $\omega_{B,C}$ to the lower transitions
\item if $\chi<0$, $\omega_{A,D}$ correspond to the lower transitions
  and $\omega_{B,C}$ to the upper transitions
\end{itemize}
Naturally, strong interactions produce splittings between the peaks
corresponding to upper and lower transitions, which may turn the
spectral doublets into fully formed quadruplets.

A second element that may be important in the system is pure
dephasing. It may be caused, depending on the physical system, by
interaction with phonons or other external elements. This effect has
proved to be relevant in state of the art semiconductor
experiments~\cite{laucht09b}. Noting the rates of dephasing of each
qubit as~$\delta_1$, $\delta_2$, one must solve the following extended
master equation from Eq.~(\ref{eq:ME}):
\begin{equation}
  \label{eq:FriJan8101404GMT2010}
  \frac{d\rho}{dt}=\mathcal{L}\rho+\sum_{i=1,2}\frac{\delta_i}{2}(2\ud{\sigma}_i\sigma_i\rho\ud{\sigma}_i\sigma_i-\ud{\sigma}_i\sigma_i\rho-\rho\ud{\sigma}_i\sigma_i)\,.
\end{equation}
In the formalism, as a consequence,
Eq.~(\ref{eq:FriJan8110055GMT2010}) becomes:
\begin{subequations}
  \label{eq:FriJan8110122GMT2010}
  \begin{align}
    &\Gamma_1^d=\gamma_1+P_1+\delta_1\,,\quad
    \Gamma_2^d=\gamma_2+P_2+\delta_2\,.
\end{align}
\end{subequations}
The dynamics in the presence of pure dephasing is again derived in
appendix~\ref{app:MonJul6000455GMT2009} and not discussed in the main
text for the same reason as before, that it spoils analyticity of the
results and brings little to the discussion. The main effect of pure
dephasing is to weaken the coherent coupling and is of a quantitative
nature, not relevant enough to discuss it further in the present text.

\section{Quantum regression formula and time correlators}
\label{app:MonJul6000455GMT2009}

In this appendix I obtain, thanks to the quantum regression formula,
the first order differential equations for the one and two-time
correlators required for the spectrum in
Eq.~(\ref{eq:TueMay5153242GMT2009}). The quantum regression theorem
states that, once we find the set of operators~$C_{\{\eta\}}$ and the
\emph{regression matrix}~$M$ that
satisfy~$\Tr{C_{\{\eta\}}\mathcal{L}O}=\sum_{\{\lambda\}}M_{\{\eta\lambda\}}\Tr{C_{\{\lambda\}}O}$
for a general operator $O$, then the equations of motion for the
two-time correlators ($\tau\geq 0$) read:
\begin{equation}
  \label{eq:TueMay5174356GMT2009}
  \frac{\partial}{\partial\tau}\mean{O(t)C_{\{\eta\}}(t+\tau)}=\sum_{\{\lambda\}}M_{\{\eta\lambda\}}\mean{O(t)C_{\{\lambda\}}(t+\tau)}
\end{equation}
The most general set of operators for the two coupled qubits
is~$C_{\{m,n,\mu,\nu\}}={\ud{\sigma_1}}^m\sigma_1^n{\ud{\sigma_2}}^\mu\sigma_2^\nu$,
with~$m$, $n$, $\mu$, $\nu\in\{0,1\}$. The corresponding regression
matrix~$M$ is:
\begin{subequations}
  \label{eq:TueDec23114907CET2008}
  \begin{align}
    &M_{\substack{mn\mu\nu\\mn\mu\nu}}=i\omega_1(m-n)+i\omega_2(\mu-\nu)-i\chi(m\mu-n\nu)\\
    &-\frac{\Gamma_1}2(m+n)-\frac{\Gamma_2}2(\mu+\nu)-\frac{\delta_1}{2}(m-n)^2-\frac{\delta_2}{2}(\mu-\nu)^2\,,\nonumber\\
    &M_{\substack{mn\mu\nu\\1-m,1-n,\mu\nu}}={P_1}mn-i\chi(1-m)(1-n)(\mu-\nu)\,,\\
    &M_{\substack{mn\mu\nu\\mn,1-\mu,1-\nu}}={P_2}\mu\nu-i\chi(1-\mu)(1-\nu)(m-n)\,,\\
    &M_{\substack{mn\mu\nu\\m,1-n,1-\mu,\nu}}=2ig(\nu-\mu)(1-n)(1-\mu)\,,\\
    &M_{\substack{mn\mu\nu\\1-m,n,\mu,1-\nu}}=2ig(n-\mu)(1-\nu)(1-m)\,,\\
    &M_{\substack{mn\mu\nu\\1-m,n,1-\mu,\nu}}=ig[m(1-\mu)+\mu(1-m)]\,,\\
    &M_{\substack{mn\mu\nu\\m,1-n,\mu,1-\nu}}=-ig[n(1-\nu)+\nu(1-n)]\,,
  \end{align}
\end{subequations}
and zero everywhere else. 

\begin{figure}[t]
  \centering
\includegraphics[width=\linewidth]{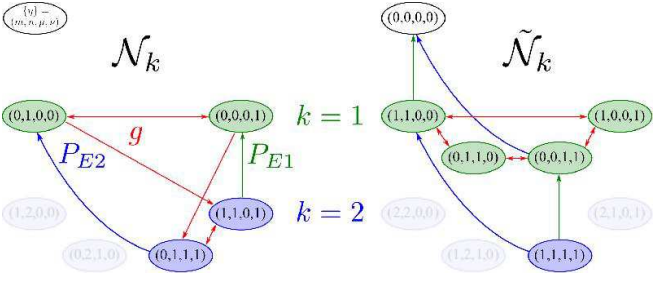}
\caption{(Color online) Chain of correlators---indexed
  by~$\{\eta\}=(m,n,\mu,\nu)$---linked by the Hamiltonian dynamics
  with pump and decay for two coupled qubits. On the left (resp.,
  right), the set~$\bigcup_k\mathcal{N}_k$ (resp.,
  $\bigcup_k\tilde{\mathcal{N}}_k$) involved in the equations of the
  two-time (resp., single-time) correlators. In green are shown the
  first manifolds~$\mathcal{N}_1$ and $\tilde{\mathcal{N}}_1$ that
  correspond to the linear model~\protect\cite{laussy09a}, and in
  blue, the second manifold~$\mathcal{N}_2$
  and~$\tilde{\mathcal{N}}_2$. The equation of motion~$\langle
  \ud{\sigma_1}(t)C_{\{\eta\}}(t+\tau)\rangle$
  with~$\eta\in\mathcal{N}_k$ requires for its initial value the
  correlator~$\langle C_{\{\tilde\eta\}}\rangle$
  with~$\{\tilde\eta\}\in\tilde{\mathcal{N}}_k$ defined
  from~$\{\eta\}=(m,n,\mu,\nu)$ by~$\{\tilde\eta\}=(m+1,n,\mu,\nu)$,
  as seen on the diagram. The thick red arrows indicate which elements
  are linked by the coherent (SC) dynamics, through the coupling
  strength~$g$, while the green/blue thin arrows show the connections
  due to the incoherent quantum dot pumpings. The sense of the arrows
  indicates which element is ``calling'' which in its equations. The
  self-coupling of each node to itself is not shown. This is
  where~$\omega_{1,2}$ and~$\Gamma_{1,2}$ enter. These links are
  obtained from the rules in Eqs.~(\ref{eq:TueDec23114907CET2008}),
  that result in the matrices~$\mathbf{M}_1$ and~$\mathbf{M}_0$. The
  number of correlators needed to compute the spectrum is truncated
  naturally (with four elements) thanks to the saturation of the
  qubit.}
\label{fig:SunJan25122441GMT2009}
\end{figure}

We concentrate on computing~$G_1^{(1)}$, as~$G_2^{(1)}$ can be
obtained from it by exchanging the indexes~$1\leftrightarrow 2$. It
corresponds to setting~$O=\ud{\sigma_1}$ and
having~$\{m,n,\mu,\nu\}=\{0,1,0,0\}$ in
Eq.~(\ref{eq:TueMay5174356GMT2009}). Other two-time correlators are
linked to the one of interest by the regression matrix, and therefore
are also needed to compute the spectrum. All correlators can be
grouped in \emph{manifolds} that, following
Refs~\cite{laussy09a,delvalle09a,delvalle09b}, I
denote~$\mathcal{N}_k$, where $k$ is the minimum number of particles
that allows the correlators to take a nonzero value. As we can see
schematically in the left part of
Fig.~\ref{fig:SunJan25122441GMT2009}, the first manifold
$\mathcal{N}_1$ is composed of the correlators with~$\{0,1,0,0\}$
and~$\{0,0,0,1\}$, linked only by the coherent coupling (red
arrows). In the linear regime (very low pump), it is enough to solve
the equations truncating in this manifold, which is the same as in the
linear model and includes~$G_1^{(1)}(\tau)$. The next (and last)
manifold~$\mathcal{N}_2$, where the two models differ, is composed of
the correlators with~$\{1,1,0,1\}$ and~$\{0,1,1,1\}$. The links
between manifolds are also of an incoherent nature, due to the pump
(green and blue arrows).

Gathering the four correlators in a vector
\begin{equation}
  \label{eq:TueDec23120004CET2008}
  \mathbf{v}(t,t+\tau)=
  \begin{pmatrix}
    \langle\ud{\sigma_1}(t)\sigma_1(t+\tau)\rangle\\
    \langle\ud{\sigma_1}(t)\sigma_2(t+\tau)\rangle\\
    \langle\ud{\sigma_1}(t)\ud{\sigma_1}\sigma_1\sigma_2(t+\tau)\rangle\\
    \langle\ud{\sigma_1}(t)\sigma_1\ud{\sigma_2}\sigma_2(t+\tau)\rangle
\end{pmatrix}
\,,
\end{equation}
their equations of motion read in matricial form:
\begin{equation}
  \label{eq:TueDec23115950CET2008}
  \frac{\partial}{\partial\tau}\mathbf{v}(t,t+\tau)=-\mathbf{M}_1\mathbf{v}(t,t+\tau)\,,
\end{equation}
with
\begin{subequations}
  \label{eq:TueDec23120838CET2008}
  \begin{align}
    &\mathbf{M}_1=\\
    &\begin{pmatrix}
      i\omega_1+\frac{\Gamma_1^\mathrm{d}}{2} & ig &-2ig&-i\chi\\
      ig & i\omega_{2}+\frac{\Gamma_2^\mathrm{d}}{2}&-i\chi&-2ig\\
      0&-P_1&i(\omega_2-\chi)+\frac{2\Gamma_1+\Gamma_2^\mathrm{d}}{2}&-ig\\
      -P_2&0&-ig&i(\omega_1-\chi)+\frac{2\Gamma_2+\Gamma_1^\mathrm{d}}{2}
\end{pmatrix}\,.\nonumber
\end{align}
\end{subequations}
The general solution (for $\tau\geq 0$),
$\mathbf{v}(t,t+\tau)=e^{-\mathbf{M}_1\tau}\mathbf{v}(t,t)$, leads to
the correlator in Eq.~(\ref{eq:TueDec23152440CET2008}) with only four
contributions, indexed by~$p=A$, $B$, $C$ and $D$. The
frequencies~$\omega_p$ and rates~$\gamma_p$ are the imaginary and real
parts of the eigenvalues of the matrix~$\mathbf{M}_1$. The
dimensionless coefficients~$L_p(t)$ and~$K_p(t)$, also real, are
linear combinations of the single-time average quantities,
\begin{equation}
\label{eq:WedJan21174259GMT2009}
\mathbf{v}(t,t)=
  \begin{pmatrix}
    \mean{\ud{\sigma_1}\sigma_1}(t)\\
    \mean{\ud{\sigma_1}\sigma_2}(t)\\
    \mean{\ud{\sigma_1}\ud{\sigma_1}\sigma_1\sigma_2}(t)\\
    \mean{\ud{\sigma_1}\sigma_1\ud{\sigma_2}\sigma_2}(t)
  \end{pmatrix}
  \,.
\end{equation}
These mean values can also be found through the quantum regression
formula, setting the operator~$O=1$ and deriving the corresponding
regression matrix from the rules in
Eq.~(\ref{eq:TueDec23114907CET2008}). They can also be grouped in
manifolds, denoted~$\tilde{\mathcal{N}}_k$. All nonzero
single-time average quantities, all of them needed to compute the
spectrum, are shown in the right part of
Fig.~\ref{fig:SunJan25122441GMT2009}. Note that
$\mean{\ud{\sigma_1}\ud{\sigma_1}\sigma_1\sigma_2}(t)=0$. The
correlators in $\tilde{\mathcal{N}}_1$ can be obtained independently
solving a second matricial equation:
\begin{equation}
  \label{eq:M6}
  \frac{d\mathbf{u}(t)}{dt}=-\mathbf{M}_0\mathbf{u}(t)+\mathbf{p}
\end{equation}
with
\begin{subequations}
    \label{eq:WedDec17235822CET2008}
  \begin{gather}
    \label{eq:WedJul8134007GMT2009}
    \mathbf{u}(t)=
    \begin{pmatrix}
      \mean{\ud{\sigma_1}\sigma_1}(t)\\
      \mean{\ud{\sigma_2}\sigma_2}(t)\\
      \mean{\ud{\sigma_1}\sigma_2}(t)\\
      \mean{\ud{\sigma_2}\sigma_1}(t)
    \end{pmatrix}
    ,\quad \mathbf{p}=
    \begin{pmatrix}
      P_1\\
      P_2\\
      0\\
      0
    \end{pmatrix}\,,
    \\[.1cm]
    \label{eq:WedJul8133652GMT2009}\mathbf{M}_0=
    \begin{pmatrix}
      \Gamma_1 & 0 & ig & -ig\\
      0 & \Gamma_2 & -ig & ig \\
      ig & -ig & \frac{\Gamma_1^\mathrm{d}+\Gamma_2^\mathrm{d}}{2}-i\Delta & 0\\
      -ig & ig & 0 & \frac{\Gamma_1^\mathrm{d}+\Gamma_2^\mathrm{d}}{2}+
      i\Delta
    \end{pmatrix}\,.
  \end{gather}
\end{subequations}
This equation is similar to that of the linear model, only changing
the effective fermionic rates into bosonic ones~$\Gamma_i\rightarrow
\tilde \Gamma_i$ \cite{laussy09a}. On the other hand, the correlator
with~$\{1,1,1,1\}$ (that is,
$\mean{\ud{\sigma_1}\sigma_1\ud{\sigma_2}\sigma_2}(t)$), which is the
only one that is not zero in~$\tilde{\mathcal{N}}_2$, finds its
expression separately, only in terms of the operators labeled
$\{1,1,0,0\}$ and $\{0,0,1,1\}$, through the equation
\begin{eqnarray}
  \label{eq:FriJan30183304CET2009}
  \frac{d}{dt}\mean{\ud{\sigma_1}\sigma_1\ud{\sigma_2}\sigma_2}(t)=-4\Gamma_+\mean{\ud{\sigma_1}\sigma_1\ud{\sigma_2}\sigma_2}(t)\nonumber\\
  +P_2\mean{\ud{\sigma_1}\sigma_1}(t)+P_1\mean{\ud{\sigma_2}\sigma_2}(t)\,.
\end{eqnarray}
In order to insert these averages in the
expression~(\ref{eq:TueMay5153242GMT2009}) for the spectrum, they must
be either time integrated, in the spontaneous emission case (SE), to
give $\mathbf{v}^\mathrm{SE}=\int_0^\infty \mathbf{v}(t,t)dt$ (and the
coefficients~$L_p^\mathrm{SE}$ and~$K_p^\mathrm{SE}$) or computed
directly in the steady state (SS) to give
$\mathbf{v}^\mathrm{SS}=\lim_{t\rightarrow\infty} \mathbf{v}(t,t)$
(and the coefficients $L_p^\mathrm{SS}$, $K_p^\mathrm{SS}$).

\subsection{Two fermions}

The regression matrix $\mathbf{M}_1$ (cf
Eq.~(\ref{eq:TueDec23120838CET2008})) truncates to its first
submatrix,
\begin{equation}
  \label{eq:SunNov15192855GMT2009}
  \mathbf{M}_1=
  \begin{pmatrix}
    i\omega_1+\frac{\Gamma_1^d}{2} & ig \\
    ig & i\omega_{2}+\frac{\Gamma_2^d}{2}
  \end{pmatrix}\,.
\end{equation}
recovering the exact mapping to the linear model, as discussed in the
text.

\subsection{Matrix of coherences}
\label{app:SatJan9144536GMT2010}

The matrix of coherence in Eq.~(\ref{eq:SatJan9144935GMT2010}) can be
found by rewriting the regression matrix $\mathbf{M}_1$ in the basis
of projectors for each transition, presented in
Eq.~(\ref{eq:MonNov16165000GMT2009}):
$\mathbf{M}_\mathrm{coh}=\mathbf{A}^{-1}\mathbf{M}_1\mathbf{A}$ with
$\mathbf{A}$ some transformation matrix. For this reason, they have
the same eigenvalues, which are the resonances of the systems. Since
the eigenstates of $\mathbf{M}_\mathrm{coh}$ are the transitions
between dressed states, their study enables us to reconstruct them
even in a highly dissipative environment.

\section{Cross Lindblad terms}
\label{app:WedJul8111440GMT2009}

In the case where the four levels of the system do not correspond to
two qubits but are the levels of a single
entity~\cite{delvalle10a}, the corresponding Liouvillian of the
system is most conveniently written in terms of the four-level
operators in Eq.~(\ref{eq:ThuOct15135642GMT2009}). The final master
equation in that case can be found from $\mathcal{L}\rho$ in
Eq.~(\ref{eq:ME}) by removing cross terms that appear in the two
coupled qubits:
\begin{align}
  \label{eq:WedJul8112044GMT2009}
  \frac{d\rho}{dt}=\mathcal{L}\rho-&\gamma_1(l_1\rho \ud{u_1}+u_1 \rho \ud{l_1})-\gamma_2(l_2\rho \ud{u_2}+u_2 \rho \ud{l_2})\\
  -&P_1(\ud{u_1}\rho l_1+\ud{l_1}\rho u_1)-P_2(\ud{u_2}\rho l_2+\ud{l_2}\rho u_2)\\
  -&\delta_1(\ud{l_1}l_1\rho \ud{u_1}u_1+ \ud{u_1}u_1 \rho \ud{l_1}l_1 )\\
  -&\delta_2(\ud{l_2}l_2\rho \ud{u_2}u_2+ \ud{u_2}u_2 \rho \ud{l_2}l_2 )\,.
\end{align}
Such cross terms represent effective couplings between the coherences,
in the case of pump and decay, and the states, in the case of pure
dephasing, associated to each qubit. Evidently, they can only have
consequences on the dynamics when the two qubits are also directly
coupled. The rules to add to those in
Eq.~(\ref{eq:TueDec23114907CET2008}) in order to obtain a new general
regression matrix~$M+\delta M$ are:
\begin{subequations}
  \label{eq:WedJul8132036GMT2009}
  \begin{align}
    &\delta M_{\substack{mn\mu\nu\\mn\mu\nu}}=P_1(m+n-1)(\mu-\nu)^2+P_2(\mu+\nu-1)(m-m)^2\,,\\
    &\delta M_{\substack{mn\mu\nu\\1-m,1-n,\mu\nu}}=-(\mu-\nu)^2[(P_1+\delta_1)mn -\tilde\Gamma_1(1-m)(1-n)]\,,\\
    &\delta M_{\substack{mn\mu\nu\\mn,1-\mu,1-\nu}}=-(m-n)^2[(P_2+\delta_2)\mu\nu-\tilde\Gamma_2(1-\mu)(1-\nu)]\,.
  \end{align}
\end{subequations}
With these new rules, the matrix for the one-time correlators,
$\mathbf{M}_0$, is still the same as in
Eq.~(\ref{eq:WedJul8133652GMT2009}) and, therefore, the mean values in
$\mathbf{u}(t)$, Eq.~(\ref{eq:WedJul8134007GMT2009}), do not differ
from those for two qubits.  The two-time regression
matrix~$\mathbf{M}_1$ changes into $\mathbf{M}_1+\delta \mathbf{M}_1$
with
\begin{subequations}
  \label{eq:WedJul8160119GMT2009}
  \begin{align}
    \delta \mathbf{M}_1=
    \begin{pmatrix}
      P_2 & 0 & 0 & \gamma_2-P_2\\
      0 & P_1 & \gamma_1-P_1 & 0\\
      0 & \delta_1+P_1 & -P_1 & 0\\
      \delta_2+P_2 & 0 & 0 & -P_2
\end{pmatrix}\,.\nonumber
\end{align}
\end{subequations}

\section{Dressed states in the manifold picture}
\label{app:SatJan9154344GMT2010}

We can give an alternative derivation and intuitive interpretation of
the results in section~\ref{sec:SunJan10204807GMT2010} by assuming
complex frequencies
$\omega_{1,2}\rightarrow\Omega_{1,2}=\omega_{1,2}-i\gamma_{1,2}/2$ in
Eq.~(\ref{eq:TueMay5132743GMT2009}). This leads to the complex dressed
frequencies:
\begin{subequations}
  \label{eq:ThuJan22112736GMT2009}
  \begin{align}
    &\omega_{\substack{+\\-}}\rightarrow\Omega_{\substack{+\\-}}=\omega_1-i\gamma_+\pm R_0\,,\\
    &\omega_{11}\rightarrow\Omega_{11}=2\omega_1-2i\gamma_+\,.
  \end{align}
\end{subequations}
From here, we can compute positions and broadenings of the four
possible contributions to the spectra. The positions are given by the
difference in energy between the levels involved in the transitions
while the broadenings are given by the sum of the imaginary parts (the
uncertainties). If we apply this principle between the levels in
Eq.~(\ref{eq:ThuJan22112736GMT2009}), we find the same result than in
Eq.~(\ref{eq:ThuDec25175033CET2008})--(\ref{eq:MonJan26115601GMT2009}). This
allows us to identify the contributions $A$, $B$, $C$, $D$ to the SC
spectrum with the manifold transitions in the order that we now
explain. In Fig.~\ref{fig:FriJan23211556GMT2009}(b) we can follow the
four possible transitions in the manifold picture. The lower
transitions (in blue), from $\ket{\pm}$, coincide, respectively, with
the expressions for peaks $A$ and $D$, given by
Eq.~(\ref{eq:ThuDec25175033CET2008}). The upper transitions (in red),
towards $\ket{\pm}$, coincide with the expressions for peaks $B$ and
$C$, given by Eq.~(\ref{eq:ThuDec25175033CET2008}). The upper
transitions have a larger broadening than the lower due to the
addition of the uncertainties in energy of the levels involved,
brought by the spontaneous decay.

In order to understand better some features of the spectra of the
steady state under incoherent pump in the different regions that we
have defined in section~\ref{sec:FriFeb6020814CET2009}, we will now
attempt to push the manifold method---adequate for vanishing pump---a
bit further, to the case of non-negligible pump. If we combine their
effects in the non-hermitian Hamiltonian as
\begin{subequations}
  \label{eq:SatJan24134658GMT2009}
  \begin{align}
    &\Omega_{0,0}=-i\frac{P_1+P_2}{2}\,,\\
    &\Omega_\pm=\omega_1-i\Gamma_+\pm
    R^\mathrm{1TD}\,,\\
    &\Omega_{11}=2\omega_1-i\frac{\gamma_1+\gamma_2}{2}\,,
  \end{align}
\end{subequations}
and apply their sum and subtraction to obtain positions and
broadenings of the spectral contributions of each of the four
transitions:
\begin{subequations}
  \label{eq:ThuAug6175705GMT2009}
  \begin{align}
    &\frac{\gamma_A}{2}+i\omega_A=\frac{3(P_1+P_2)+\gamma_1+\gamma_2}{4}+iR^\mathrm{1TD}\,,\\
    &\frac{\gamma_B}{2}+i\omega_B=\frac{3(\gamma_1+\gamma_2)+P_1+P_2}{4}+iR^\mathrm{1TD}\,,\\
    &\frac{\gamma_C}{2}+i\omega_C=\frac{3(\gamma_1+\gamma_2)+P_1+P_2}{4}-iR^\mathrm{1TD}\,,\\
    &\frac{\gamma_D}{2}+i\omega_D=\frac{3(P_1+P_2)+\gamma_1+\gamma_2}{4}-iR^\mathrm{1TD}\,,
  \end{align}
\end{subequations}
we do not obtain the right results of four---in principle
different---peaks. This discussion brings us back to the naive
association of SC with $\Re(R^\mathrm{1TD})\neq 0$ we discarded in
Sec.~\ref{sec:SunFeb1051813CET2009}. It is evident that using only the
single-time dynamics frequency $R^\mathrm{1TD}$ to describe the
splitting between the dressed states, cannot give the variety of
situations than the two complex parameters~$z_{1,2}$ give.

In the case of a 4LS, the results obtained in
Eq.~(\ref{eq:WedAug5140534GMT2009}) are comparable to those in
Eq.~(\ref{eq:ThuAug6175705GMT2009}), even though the positions of the
peaks are given by $\tilde R$ instead of $R^\mathrm{1TD}$. The
broadenings are exactly the same. We can conclude that without the
cross Lindblad terms, the coupling acquires a bosonic character: the
correlations between levels are relaxed from those of two qubits into
those of one (larger) four-level system where the manifold picture
holds.

\section{Second order correlations functions}
\label{app:SunFeb1121831CET2009}

In this appendix, I obtain the second order direct and cross
correlation functions defined, respectively, as
\begin{subequations}
  \label{eq:ThuMay7122454GMT2009}
  \begin{align}
    &G_1^{(2)}(t,t+\tau)=\mean{\ud{\sigma_1}(t)\ud{\sigma_1}(t+\tau)\sigma_1(t+\tau)\sigma_1(t)}\,,\\
    &G_{12}^{(2)}(t,t+\tau)=\mean{\ud{\sigma_1}(t)\ud{\sigma_2}(t+\tau)\sigma_2(t+\tau)\sigma_1(t)}\,.
\end{align}
\end{subequations}
These two quantities are proportional to the probabilities of two
$\tau$-delayed emissions from the same qubit ($G_1^{(2)}$) or two
different qubit emissions ($G_{12}^{(2)}$).

\begin{figure}[h]
  \centering
  \includegraphics[width=0.45\linewidth]{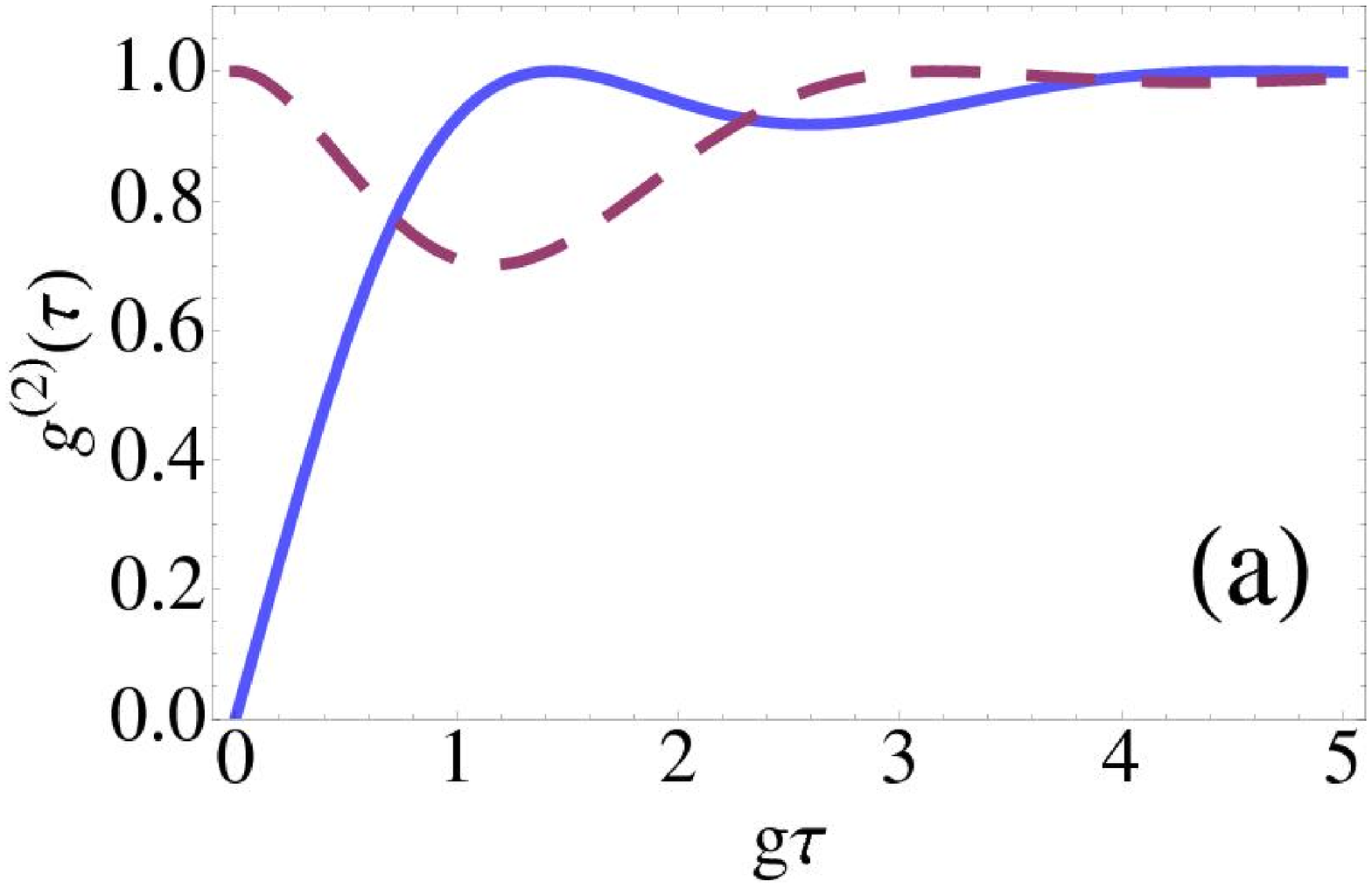}
  \includegraphics[width=0.45\linewidth]{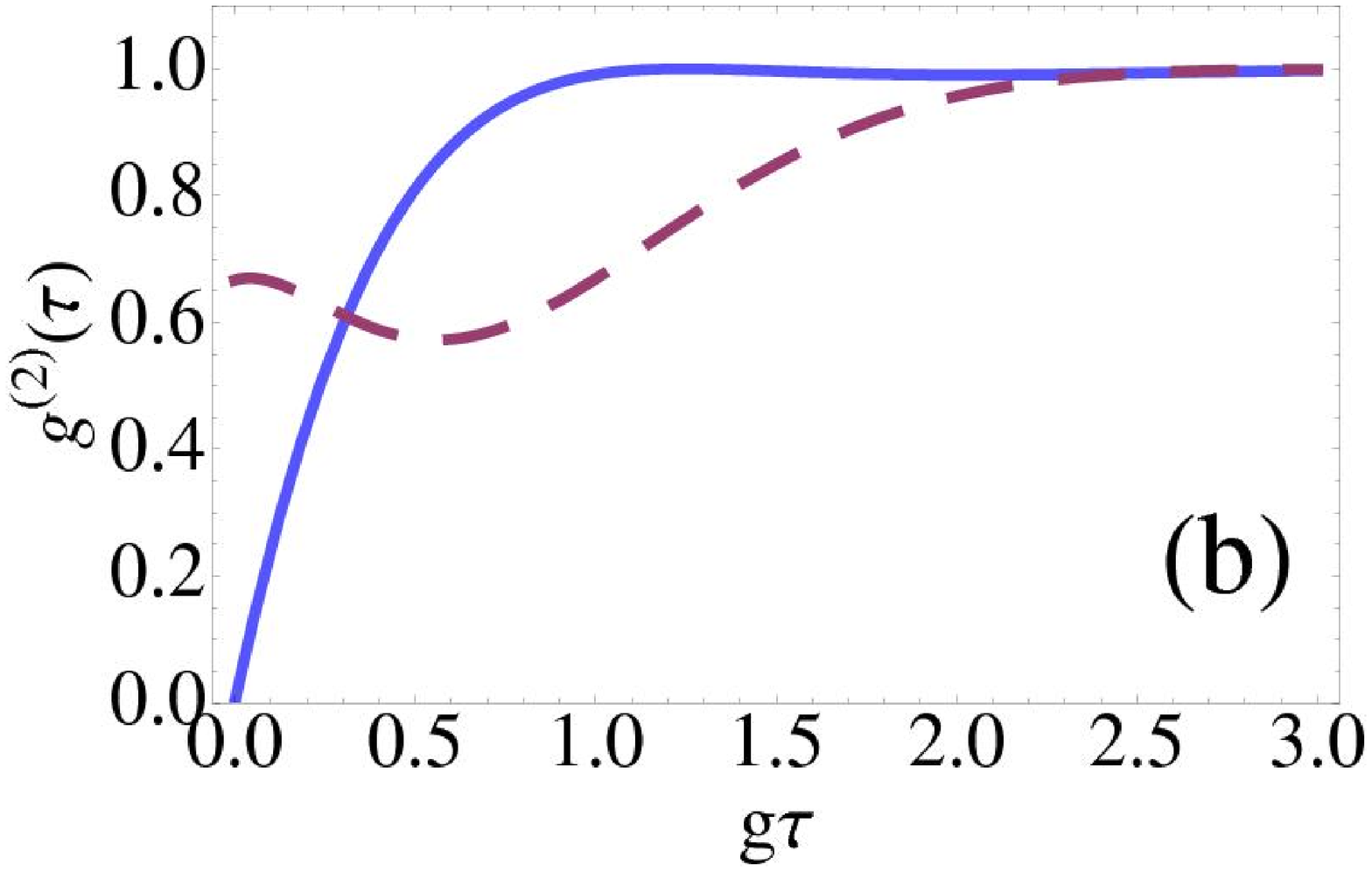}
  \caption{(Color online) Steady state values of $g^{(2)}(\tau)$
    (solid blue) $g_{12}^{(2)}(\tau)$ (dashed purple) at
    resonance. (a) The system is in FSC: $\gamma_1=g$, $\gamma_2=g/2$,
    $P_1=g/2$ and $P_2=g/10$. (b) In MC with optimum coupling
    $G=\sqrt{2}g$: $\gamma_1=P_2=0$ and $P_1=\gamma_2=2g$ (see
    Fig.~\ref{fig:SunFeb1041856CET2009}).}
\label{fig:FriFeb13203447CET2009}
\end{figure}
In order to compute such two-time correlators, we apply the quantum
regression formula presented in an alternative form to
Eq.~(\ref{eq:TueMay5174356GMT2009}):
\begin{equation}
  \label{eq:ThuMay7122942GMT2009}
  \frac{\partial}{\partial\tau}\mean{O(t)C_{\{\eta\}}(t+\tau)Q(t)}=\sum_{\{\lambda\}}M_{\{\eta\lambda\}}\mean{O(t)C_{\{\lambda\}}(t+\tau)Q(t)}
\end{equation}
where $Q$ is a second arbitrary operator. The equations we must solve
setting $O=\ud{\sigma_1}$ and $Q=\sigma_1$, are the same as those used
to compute the mean values in Eq.~(\ref{eq:M6}) but now with the
vector of correlators
\begin{equation}
  \label{eq:TueDec23120004CET20082}
  \mathbf{w}(t,t+\tau)=
  \begin{pmatrix}
    \mean{\ud{\sigma_1}(t)\ud{\sigma_1}\sigma_1(t+\tau)\sigma_1(t)}\\
    \mean{\ud{\sigma_1}(t)\ud{\sigma_2}\sigma_2(t+\tau)\sigma_1(t)}\\
    \mean{\ud{\sigma_1}(t)\ud{\sigma_1}\sigma_2(t+\tau)\sigma_1(t)}\\
    \mean{\ud{\sigma_1}(t)\sigma_1\ud{\sigma_2}(t+\tau)\sigma_1(t)}
\end{pmatrix}
\,,
\end{equation}
and the vector $n_1(t)\mathbf{p}$ instead of $\mathbf{p}$. In the
steady state ($t=0$), the solution for $\mathbf{w}(0,\tau)$ requires
knowing the initial condition:
\begin{eqnarray}
  \label{eq:SunFeb1124957CET2009}
  \mathbf{w}^\mathrm{SS}=\mathbf{w}(0,0)=
  \left(
\begin{array}{c}
  0\\
  n_{11}\\
  0\\
  0
\end{array}\right)\,.
\end{eqnarray}
In terms of the following vector of mean fluctuations,
\begin{eqnarray}
  \label{eq:ThuMay7124624GMT2009}
  \mathbf{f}^\mathrm{SS}=-(\mathbf{w}^\mathrm{SS}-\mathbf{u}^\mathrm{SS}n_1^\mathrm{SS})=
  \left(
\begin{array}{c}
  n_1^2\\
  n_1n_2-n_{11}\\
  n_\mathrm{corr}n_1\\
  n_\mathrm{corr}^*n_1
\end{array}\right)\,,
\end{eqnarray}
the steady state normalized correlations functions (for $\tau>0$) read
\begin{subequations}
  \label{eq:SunFeb1125151CET2009}
  \begin{align}
    &g^{(2)}(\tau)=\frac{G_1^{(2)}(0,\tau)}{n_1^2}=1-\frac{[e^{-\mathbf{M}_0\tau}\,\mathbf{f}^\mathrm{SS}]_1}{n_1^2}\,\in[0,1],\\
    &g_{12}^{(2)}(\tau)=\frac{G_{12}^{(2)}(0,\tau)}{n_1n_2}=1-\frac{[e^{-\mathbf{M}_0\tau}\,\mathbf{f}^\mathrm{SS}]_2}{n_1n_2}\,\in[0,1]
  \end{align}
\end{subequations}
where $[\mathbf{x}]_1$ means that we take the first element of the
vector~$\mathbf{x}$. At infinite delays, the two functions reach the
value of uncorrelated emissions,
\begin{equation}
    \label{eq:ThuMay7125847GMT2009}
    g_1^{(2)}(\tau\rightarrow\infty)\,,\,g_{12}^{(2)}(\tau\rightarrow\infty)\rightarrow 1\,.
\end{equation}

At zero delay, it is evident that $g^{(2)}(0)=0$, as two excitations
cannot exist in the same qubit. At intermediate delays, the emission
presents \emph{antibunching}: $g^{(2)}(0)<g^{(2)}(\tau)$, the second
excitation from the same qubit has more probability to be emitted
after some delay (see Fig.~\ref{fig:FriFeb13203447CET2009}). At zero
delay,
\begin{equation}
  \label{eq:ThuMay7143451GMT2009}
  g_{12}^{(2)}(0)=\frac{n_{11}}{n_1n_2}\,.
\end{equation}
It always is possible a second emission coming from a different qubit
(the minimum value for $g_{12}^{(2)}(\tau=0)$ is $1/5$). When the qubit
behave as independently, $g_{12}^{(2)}(\tau)=1$ all along. 

Being the second order correlation functions given by basically the
same equation as the populations and depending only on them,
$g_1^{(2)}(\tau)$ and $g_{12}^{(2)}(\tau)$ are the same for the two
coupled fermions, qubits and 4LS. It is not possible to distinguish
from their dynamics the four coupling regions described in
Table~\ref{tab:SatJan9121355GMT2010}, appearing in the case of coupled
qubits.

\bibliography{Sci,books}

\begin{thebibliography}{38}
\expandafter\ifx\csname natexlab\endcsname\relax\def\natexlab#1{#1}\fi
\expandafter\ifx\csname bibnamefont\endcsname\relax
  \def\bibnamefont#1{#1}\fi
\expandafter\ifx\csname bibfnamefont\endcsname\relax
  \def\bibfnamefont#1{#1}\fi
\expandafter\ifx\csname citenamefont\endcsname\relax
  \def\citenamefont#1{#1}\fi
\expandafter\ifx\csname url\endcsname\relax
  \def\url#1{\texttt{#1}}\fi
\expandafter\ifx\csname urlprefix\endcsname\relax\def\urlprefix{URL }\fi
\providecommand{\bibinfo}[2]{#2}
\providecommand{\eprint}[2][]{\url{#2}}

\bibitem[{\citenamefont{Nielsen and Chuang}(2000)}]{nielsen_book00a}
\bibinfo{author}{\bibfnamefont{M.~A.} \bibnamefont{Nielsen}} \bibnamefont{and}
  \bibinfo{author}{\bibfnamefont{I.~L.} \bibnamefont{Chuang}},
  \emph{\bibinfo{title}{Quantum computation and quantum information}}
  (\bibinfo{publisher}{Cambridge University Press}, \bibinfo{year}{2000}).

\bibitem[{\citenamefont{Gardiner and Zoller}(2000)}]{gardiner_book00a}
\bibinfo{author}{\bibfnamefont{G.~W.} \bibnamefont{Gardiner}} \bibnamefont{and}
  \bibinfo{author}{\bibfnamefont{P.}~\bibnamefont{Zoller}},
  \emph{\bibinfo{title}{Quantum Noise}} (\bibinfo{publisher}{Springer-Verlag,
  Berlin}, \bibinfo{year}{2000}), \bibinfo{edition}{2nd} ed.

\bibitem[{\citenamefont{Haroche and Raimond}(2006)}]{haroche_book06a}
\bibinfo{author}{\bibfnamefont{S.}~\bibnamefont{Haroche}} \bibnamefont{and}
  \bibinfo{author}{\bibfnamefont{J.-M.} \bibnamefont{Raimond}},
  \emph{\bibinfo{title}{Exploring the Quantum: Atoms, Cavities, and Photons}}
  (\bibinfo{publisher}{Oxford University Press}, \bibinfo{year}{2006}).

\bibitem[{\citenamefont{Allen and Eberly}(1987)}]{allen_book87a}
\bibinfo{author}{\bibfnamefont{L.}~\bibnamefont{Allen}} \bibnamefont{and}
  \bibinfo{author}{\bibfnamefont{J.~H.} \bibnamefont{Eberly}},
  \emph{\bibinfo{title}{Optical Resonance and Two-Level Atoms}}
  (\bibinfo{publisher}{Dover}, \bibinfo{year}{1987}).

\bibitem[{\citenamefont{Kavokin et~al.}(2007)\citenamefont{Kavokin, Baumberg,
  Malpuech, and Laussy}}]{kavokin_book07a}
\bibinfo{author}{\bibfnamefont{A.}~\bibnamefont{Kavokin}},
  \bibinfo{author}{\bibfnamefont{J.~J.} \bibnamefont{Baumberg}},
  \bibinfo{author}{\bibfnamefont{G.}~\bibnamefont{Malpuech}}, \bibnamefont{and}
  \bibinfo{author}{\bibfnamefont{F.~P.} \bibnamefont{Laussy}},
  \emph{\bibinfo{title}{Microcavities}} (\bibinfo{publisher}{Oxford University
  Press}, \bibinfo{year}{2007}).

\bibitem[{\citenamefont{Cohen-Tannoudji
  et~al.}(2001)\citenamefont{Cohen-Tannoudji, Dupont-Roc, and
  Grynberg}}]{cohentannoudji_book01a}
\bibinfo{author}{\bibfnamefont{C.}~\bibnamefont{Cohen-Tannoudji}},
  \bibinfo{author}{\bibfnamefont{J.}~\bibnamefont{Dupont-Roc}},
  \bibnamefont{and} \bibinfo{author}{\bibfnamefont{G.}~\bibnamefont{Grynberg}},
  \emph{\bibinfo{title}{Photons et atomes}} (\bibinfo{publisher}{EDP Sciences},
  \bibinfo{year}{2001}).

\bibitem[{\citenamefont{Sanchez-Mondragon
  et~al.}(1983)\citenamefont{Sanchez-Mondragon, Narozhny, and
  Eberly}}]{sanchezmondragon83a}
\bibinfo{author}{\bibfnamefont{J.~J.} \bibnamefont{Sanchez-Mondragon}},
  \bibinfo{author}{\bibfnamefont{N.~B.} \bibnamefont{Narozhny}},
  \bibnamefont{and} \bibinfo{author}{\bibfnamefont{J.~H.}
  \bibnamefont{Eberly}}, \bibinfo{journal}{Phys. Rev. Lett.}
  \textbf{\bibinfo{volume}{51}}, \bibinfo{pages}{550} (\bibinfo{year}{1983}).

\bibitem[{\citenamefont{Carmichael et~al.}(1989)\citenamefont{Carmichael,
  Brecha, Raizen, Kimble, and Rice}}]{carmichael89a}
\bibinfo{author}{\bibfnamefont{H.~J.} \bibnamefont{Carmichael}},
  \bibinfo{author}{\bibfnamefont{R.~J.} \bibnamefont{Brecha}},
  \bibinfo{author}{\bibfnamefont{M.~G.} \bibnamefont{Raizen}},
  \bibinfo{author}{\bibfnamefont{H.~J.} \bibnamefont{Kimble}},
  \bibnamefont{and} \bibinfo{author}{\bibfnamefont{P.~R.} \bibnamefont{Rice}},
  \bibinfo{journal}{Phys. Rev. A} \textbf{\bibinfo{volume}{40}},
  \bibinfo{pages}{5516} (\bibinfo{year}{1989}).

\bibitem[{\citenamefont{Mollow}(1969)}]{mollow69a}
\bibinfo{author}{\bibfnamefont{B.~R.} \bibnamefont{Mollow}},
  \bibinfo{journal}{Phys. Rev.} \textbf{\bibinfo{volume}{188}},
  \bibinfo{pages}{1969} (\bibinfo{year}{1969}).

\bibitem[{\citenamefont{Cohen-Tannoudji and Reynaud}(1977)}]{cohentannoudji77a}
\bibinfo{author}{\bibfnamefont{C.}~\bibnamefont{Cohen-Tannoudji}}
  \bibnamefont{and} \bibinfo{author}{\bibfnamefont{S.}~\bibnamefont{Reynaud}},
  \bibinfo{journal}{J. phys. B.: At. Mol. Phys.} \textbf{\bibinfo{volume}{10}},
  \bibinfo{pages}{345} (\bibinfo{year}{1977}).

\bibitem[{\citenamefont{Cirac et~al.}(1991)\citenamefont{Cirac, Ritsch, and
  Zoller}}]{cirac91a}
\bibinfo{author}{\bibfnamefont{J.~I.} \bibnamefont{Cirac}},
  \bibinfo{author}{\bibfnamefont{H.}~\bibnamefont{Ritsch}}, \bibnamefont{and}
  \bibinfo{author}{\bibfnamefont{P.}~\bibnamefont{Zoller}},
  \bibinfo{journal}{Phys. Rev. A} \textbf{\bibinfo{volume}{44}},
  \bibinfo{pages}{4541} (\bibinfo{year}{1991}).

\bibitem[{\citenamefont{Tian and Carmichael}(1992)}]{tian92a}
\bibinfo{author}{\bibfnamefont{L.}~\bibnamefont{Tian}} \bibnamefont{and}
  \bibinfo{author}{\bibfnamefont{H.~J.} \bibnamefont{Carmichael}},
  \bibinfo{journal}{Quantum Opt.} \textbf{\bibinfo{volume}{4}},
  \bibinfo{pages}{131} (\bibinfo{year}{1992}).

\bibitem[{\citenamefont{del Valle}(2009)}]{delvalle_book09a}
\bibinfo{author}{\bibfnamefont{E.}~\bibnamefont{del Valle}},
  \emph{\bibinfo{title}{Microcavity Quantum Electrodynamics}}
  (\bibinfo{publisher}{VDM Verlag}, \bibinfo{year}{2009}).

\bibitem[{\citenamefont{Laussy et~al.}(2009)\citenamefont{Laussy, del Valle,
  and Tejedor}}]{laussy09a}
\bibinfo{author}{\bibfnamefont{F.~P.} \bibnamefont{Laussy}},
  \bibinfo{author}{\bibfnamefont{E.}~\bibnamefont{del Valle}},
  \bibnamefont{and} \bibinfo{author}{\bibfnamefont{C.}~\bibnamefont{Tejedor}},
  \bibinfo{journal}{Phys. Rev. B} \textbf{\bibinfo{volume}{79}},
  \bibinfo{pages}{235325} (\bibinfo{year}{2009}).

\bibitem[{\citenamefont{Jaynes and Cummings}(1963)}]{jaynes63a}
\bibinfo{author}{\bibfnamefont{E.}~\bibnamefont{Jaynes}} \bibnamefont{and}
  \bibinfo{author}{\bibfnamefont{F.}~\bibnamefont{Cummings}},
  \bibinfo{journal}{Proc. IEEE} \textbf{\bibinfo{volume}{51}},
  \bibinfo{pages}{89} (\bibinfo{year}{1963}).

\bibitem[{\citenamefont{del Valle et~al.}(2009{\natexlab{a}})\citenamefont{del
  Valle, Laussy, and Tejedor}}]{delvalle09a}
\bibinfo{author}{\bibfnamefont{E.}~\bibnamefont{del Valle}},
  \bibinfo{author}{\bibfnamefont{F.~P.} \bibnamefont{Laussy}},
  \bibnamefont{and} \bibinfo{author}{\bibfnamefont{C.}~\bibnamefont{Tejedor}},
  \bibinfo{journal}{Phys. Rev. B} \textbf{\bibinfo{volume}{79}},
  \bibinfo{pages}{235326} (\bibinfo{year}{2009}{\natexlab{a}}).

\bibitem[{\citenamefont{Pashkin et~al.}(2009)\citenamefont{Pashkin, Astafiev,
  Yamamoto, Nakamura, and Tsai}}]{pashkin09a}
\bibinfo{author}{\bibfnamefont{Y.~A.} \bibnamefont{Pashkin}},
  \bibinfo{author}{\bibfnamefont{O.}~\bibnamefont{Astafiev}},
  \bibinfo{author}{\bibfnamefont{T.}~\bibnamefont{Yamamoto}},
  \bibinfo{author}{\bibfnamefont{Y.}~\bibnamefont{Nakamura}}, \bibnamefont{and}
  \bibinfo{author}{\bibfnamefont{J.~S.} \bibnamefont{Tsai}},
  \bibinfo{journal}{Quantum Information Processing}
  \textbf{\bibinfo{volume}{8}}, \bibinfo{pages}{55} (\bibinfo{year}{2009}).

\bibitem[{\citenamefont{Pashkin et~al.}(2003)\citenamefont{Pashkin, Yamamoto,
  Astafiev, Nakamura, Averin, and Tsai}}]{pashkin03a}
\bibinfo{author}{\bibfnamefont{Y.~A.} \bibnamefont{Pashkin}},
  \bibinfo{author}{\bibfnamefont{T.}~\bibnamefont{Yamamoto}},
  \bibinfo{author}{\bibfnamefont{O.}~\bibnamefont{Astafiev}},
  \bibinfo{author}{\bibfnamefont{Y.}~\bibnamefont{Nakamura}},
  \bibinfo{author}{\bibfnamefont{D.~V.} \bibnamefont{Averin}},
  \bibnamefont{and} \bibinfo{author}{\bibfnamefont{J.~S.} \bibnamefont{Tsai}},
  \bibinfo{journal}{Nature} \textbf{\bibinfo{volume}{421}},
  \bibinfo{pages}{823} (\bibinfo{year}{2003}).

\bibitem[{\citenamefont{Majer et~al.}(2005)\citenamefont{Majer, Paauw, ter
  Haar, Harmans, and Mooij}}]{majer05a}
\bibinfo{author}{\bibfnamefont{J.~B.} \bibnamefont{Majer}},
  \bibinfo{author}{\bibfnamefont{F.~G.} \bibnamefont{Paauw}},
  \bibinfo{author}{\bibfnamefont{A.~C.~J.} \bibnamefont{ter Haar}},
  \bibinfo{author}{\bibfnamefont{C.~J. P.~M.} \bibnamefont{Harmans}},
  \bibnamefont{and} \bibinfo{author}{\bibfnamefont{J.~E.} \bibnamefont{Mooij}},
  \bibinfo{journal}{Phys. Rev. Lett.} \textbf{\bibinfo{volume}{94}},
  \bibinfo{pages}{090501} (\bibinfo{year}{2005}).

\bibitem[{\citenamefont{Hime et~al.}(2006)\citenamefont{Hime, Reichardt,
  Plourde, Robertson, Wu, Ustinov, and Clarke}}]{hime06a}
\bibinfo{author}{\bibfnamefont{T.}~\bibnamefont{Hime}},
  \bibinfo{author}{\bibfnamefont{P.~A.} \bibnamefont{Reichardt}},
  \bibinfo{author}{\bibfnamefont{B.~L.~T.} \bibnamefont{Plourde}},
  \bibinfo{author}{\bibfnamefont{T.~L.} \bibnamefont{Robertson}},
  \bibinfo{author}{\bibfnamefont{C.-E.} \bibnamefont{Wu}},
  \bibinfo{author}{\bibfnamefont{A.~V.} \bibnamefont{Ustinov}},
  \bibnamefont{and} \bibinfo{author}{\bibfnamefont{J.}~\bibnamefont{Clarke}},
  \bibinfo{journal}{Science} \textbf{\bibinfo{volume}{314}},
  \bibinfo{pages}{1427} (\bibinfo{year}{2006}).

\bibitem[{\citenamefont{Niskanen et~al.}(2007)\citenamefont{Niskanen, Harrabi,
  Yoshihara, Nakamura, Lloyd, and Tsai}}]{niskanen07a}
\bibinfo{author}{\bibfnamefont{A.~O.} \bibnamefont{Niskanen}},
  \bibinfo{author}{\bibfnamefont{K.}~\bibnamefont{Harrabi}},
  \bibinfo{author}{\bibfnamefont{F.}~\bibnamefont{Yoshihara}},
  \bibinfo{author}{\bibfnamefont{Y.}~\bibnamefont{Nakamura}},
  \bibinfo{author}{\bibfnamefont{S.}~\bibnamefont{Lloyd}}, \bibnamefont{and}
  \bibinfo{author}{\bibfnamefont{J.~S.} \bibnamefont{Tsai}},
  \bibinfo{journal}{Science} \textbf{\bibinfo{volume}{316}},
  \bibinfo{pages}{723} (\bibinfo{year}{2007}).

\bibitem[{\citenamefont{Wootters}(1998)}]{wootters98a}
\bibinfo{author}{\bibfnamefont{W.~K.} \bibnamefont{Wootters}},
  \bibinfo{journal}{Phys. Rev. Lett.} \textbf{\bibinfo{volume}{80}},
  \bibinfo{pages}{2245} (\bibinfo{year}{1998}).

\bibitem[{\citenamefont{Berkley et~al.}(2003)\citenamefont{Berkley, Xu, Ramos,
  Gubrud, Strauch, Johnson, Anderson, Dragt, Lobb, and Wellstood}}]{berkley03a}
\bibinfo{author}{\bibfnamefont{A.~J.} \bibnamefont{Berkley}},
  \bibinfo{author}{\bibfnamefont{H.}~\bibnamefont{Xu}},
  \bibinfo{author}{\bibfnamefont{R.~C.} \bibnamefont{Ramos}},
  \bibinfo{author}{\bibfnamefont{M.~A.} \bibnamefont{Gubrud}},
  \bibinfo{author}{\bibfnamefont{F.~W.} \bibnamefont{Strauch}},
  \bibinfo{author}{\bibfnamefont{P.~R.} \bibnamefont{Johnson}},
  \bibinfo{author}{\bibfnamefont{J.~R.} \bibnamefont{Anderson}},
  \bibinfo{author}{\bibfnamefont{A.~J.} \bibnamefont{Dragt}},
  \bibinfo{author}{\bibfnamefont{C.~J.} \bibnamefont{Lobb}}, \bibnamefont{and}
  \bibinfo{author}{\bibfnamefont{F.~C.} \bibnamefont{Wellstood}},
  \bibinfo{journal}{Science} \textbf{\bibinfo{volume}{300}},
  \bibinfo{pages}{1548} (\bibinfo{year}{2003}).

\bibitem[{\citenamefont{Yamamoto et~al.}(2003)\citenamefont{Yamamoto, Pashkin,
  Astafiev, Nakamura, and Tsai}}]{yamamoto03a}
\bibinfo{author}{\bibfnamefont{T.}~\bibnamefont{Yamamoto}},
  \bibinfo{author}{\bibfnamefont{Y.~A.} \bibnamefont{Pashkin}},
  \bibinfo{author}{\bibfnamefont{O.}~\bibnamefont{Astafiev}},
  \bibinfo{author}{\bibfnamefont{Y.}~\bibnamefont{Nakamura}}, \bibnamefont{and}
  \bibinfo{author}{\bibfnamefont{J.~S.} \bibnamefont{Tsai}},
  \bibinfo{journal}{Nature} \textbf{\bibinfo{volume}{425}},
  \bibinfo{pages}{941} (\bibinfo{year}{2003}).

\bibitem[{\citenamefont{Li et~al.}(2003)\citenamefont{Li, Wu, Steel, Gammon,
  Stievater, Katzer, Park, Piermarocchi, and Sham}}]{li03a}
\bibinfo{author}{\bibfnamefont{X.}~\bibnamefont{Li}},
  \bibinfo{author}{\bibfnamefont{Y.}~\bibnamefont{Wu}},
  \bibinfo{author}{\bibfnamefont{D.}~\bibnamefont{Steel}},
  \bibinfo{author}{\bibfnamefont{D.}~\bibnamefont{Gammon}},
  \bibinfo{author}{\bibfnamefont{T.~H.} \bibnamefont{Stievater}},
  \bibinfo{author}{\bibfnamefont{D.~S.} \bibnamefont{Katzer}},
  \bibinfo{author}{\bibfnamefont{D.}~\bibnamefont{Park}},
  \bibinfo{author}{\bibfnamefont{C.}~\bibnamefont{Piermarocchi}},
  \bibnamefont{and} \bibinfo{author}{\bibfnamefont{L.~J.} \bibnamefont{Sham}},
  \bibinfo{journal}{Science} \textbf{\bibinfo{volume}{301}},
  \bibinfo{pages}{809} (\bibinfo{year}{2003}).

\bibitem[{\citenamefont{Grigorenko and Khveshchenko}(2005)}]{grigorenko05a}
\bibinfo{author}{\bibfnamefont{I.~A.} \bibnamefont{Grigorenko}}
  \bibnamefont{and} \bibinfo{author}{\bibfnamefont{D.~V.}
  \bibnamefont{Khveshchenko}}, \bibinfo{journal}{Phys. Rev. Lett.}
  \textbf{\bibinfo{volume}{94}}, \bibinfo{pages}{040506}
  (\bibinfo{year}{2005}).

\bibitem[{\citenamefont{Galperin et~al.}(2005)\citenamefont{Galperin, Shantsev,
  Bergli, and Altshuler}}]{galperin05a}
\bibinfo{author}{\bibfnamefont{Y.~M.} \bibnamefont{Galperin}},
  \bibinfo{author}{\bibfnamefont{D.~V.} \bibnamefont{Shantsev}},
  \bibinfo{author}{\bibfnamefont{J.}~\bibnamefont{Bergli}}, \bibnamefont{and}
  \bibinfo{author}{\bibfnamefont{B.~L.} \bibnamefont{Altshuler}},
  \bibinfo{journal}{Europhys. Lett.} \textbf{\bibinfo{volume}{71}},
  \bibinfo{pages}{21} (\bibinfo{year}{2005}).

\bibitem[{\citenamefont{Dicke}(1954)}]{dicke54a}
\bibinfo{author}{\bibfnamefont{R.~H.} \bibnamefont{Dicke}},
  \bibinfo{journal}{Phys. Rev.} \textbf{\bibinfo{volume}{93}},
  \bibinfo{pages}{99} (\bibinfo{year}{1954}).

\bibitem[{\citenamefont{Shirokov}(1990)}]{shirokov90a}
\bibinfo{author}{\bibfnamefont{M.~I.} \bibnamefont{Shirokov}},
  \bibinfo{journal}{J. phys. B.: At. Mol. Phys.} \textbf{\bibinfo{volume}{23}},
  \bibinfo{pages}{1923} (\bibinfo{year}{1990}).

\bibitem[{\citenamefont{Messiah}(1999)}]{messiah_book99a}
\bibinfo{author}{\bibfnamefont{A.}~\bibnamefont{Messiah}},
  \emph{\bibinfo{title}{Quantum Mechanics}} (\bibinfo{publisher}{Dover},
  \bibinfo{year}{1999}).

\bibitem[{\citenamefont{Morigi et~al.}(2002)\citenamefont{Morigi,
  Franke-Arnold, and Oppo}}]{morigi02a}
\bibinfo{author}{\bibfnamefont{G.}~\bibnamefont{Morigi}},
  \bibinfo{author}{\bibfnamefont{S.}~\bibnamefont{Franke-Arnold}},
  \bibnamefont{and} \bibinfo{author}{\bibfnamefont{G.-L.} \bibnamefont{Oppo}},
  \bibinfo{journal}{Phys. Rev. A} \textbf{\bibinfo{volume}{66}},
  \bibinfo{pages}{053409} (\bibinfo{year}{2002}).

\bibitem[{\citenamefont{del Valle et~al.}(2010)\citenamefont{del Valle,
  Zippilli, Laussy, Gonzalez-Tudela, Morigi, and Tejedor}}]{delvalle10a}
\bibinfo{author}{\bibfnamefont{E.}~\bibnamefont{del Valle}},
  \bibinfo{author}{\bibfnamefont{S.}~\bibnamefont{Zippilli}},
  \bibinfo{author}{\bibfnamefont{F.~P.} \bibnamefont{Laussy}},
  \bibinfo{author}{\bibfnamefont{A.}~\bibnamefont{Gonzalez-Tudela}},
  \bibinfo{author}{\bibfnamefont{G.}~\bibnamefont{Morigi}}, \bibnamefont{and}
  \bibinfo{author}{\bibfnamefont{C.}~\bibnamefont{Tejedor}},
  \bibinfo{journal}{Phys. Rev. B} \textbf{\bibinfo{volume}{81}},
  \bibinfo{pages}{035302} (\bibinfo{year}{2010}).

\bibitem[{\citenamefont{Gerardot et~al.}(2005)\citenamefont{Gerardot, Strauf,
  de~Dood, Bychkov, Badolato, Hennessy, Hu, Bouwmeester, and
  Petroff}}]{gerardot05a}
\bibinfo{author}{\bibfnamefont{B.~D.} \bibnamefont{Gerardot}},
  \bibinfo{author}{\bibfnamefont{S.}~\bibnamefont{Strauf}},
  \bibinfo{author}{\bibfnamefont{M.~J.~A.} \bibnamefont{de~Dood}},
  \bibinfo{author}{\bibfnamefont{A.~M.} \bibnamefont{Bychkov}},
  \bibinfo{author}{\bibfnamefont{A.}~\bibnamefont{Badolato}},
  \bibinfo{author}{\bibfnamefont{K.}~\bibnamefont{Hennessy}},
  \bibinfo{author}{\bibfnamefont{E.~L.} \bibnamefont{Hu}},
  \bibinfo{author}{\bibfnamefont{D.}~\bibnamefont{Bouwmeester}},
  \bibnamefont{and} \bibinfo{author}{\bibfnamefont{P.~M.}
  \bibnamefont{Petroff}}, \bibinfo{journal}{Phys. Rev. Lett.}
  \textbf{\bibinfo{volume}{95}}, \bibinfo{pages}{137403}
  (\bibinfo{year}{2005}).

\bibitem[{\citenamefont{Carmichael}(2002)}]{carmichael_book02a}
\bibinfo{author}{\bibfnamefont{H.~J.} \bibnamefont{Carmichael}},
  \emph{\bibinfo{title}{Statistical methods in quantum optics 1}}
  (\bibinfo{publisher}{Springer}, \bibinfo{year}{2002}), \bibinfo{edition}{2nd}
  ed.

\bibitem[{\citenamefont{Briegel and Englert}(1993)}]{briegel93a}
\bibinfo{author}{\bibfnamefont{H.-J.} \bibnamefont{Briegel}} \bibnamefont{and}
  \bibinfo{author}{\bibfnamefont{B.-G.} \bibnamefont{Englert}},
  \bibinfo{journal}{Phys. Rev. A} \textbf{\bibinfo{volume}{47}},
  \bibinfo{pages}{3311} (\bibinfo{year}{1993}).

\bibitem[{\citenamefont{del Valle et~al.}(2009{\natexlab{b}})\citenamefont{del
  Valle, Laussy, and Tejedor}}]{delvalle09b}
\bibinfo{author}{\bibfnamefont{E.}~\bibnamefont{del Valle}},
  \bibinfo{author}{\bibfnamefont{F.~P.} \bibnamefont{Laussy}},
  \bibnamefont{and} \bibinfo{author}{\bibfnamefont{C.}~\bibnamefont{Tejedor}},
  \bibinfo{journal}{AIP Conference Proceedings}
  \textbf{\bibinfo{volume}{1147}}, \bibinfo{pages}{238}
  (\bibinfo{year}{2009}{\natexlab{b}}).

\bibitem[{\citenamefont{Laussy et~al.}(2008)\citenamefont{Laussy, del Valle,
  and Tejedor}}]{laussy08a}
\bibinfo{author}{\bibfnamefont{F.~P.} \bibnamefont{Laussy}},
  \bibinfo{author}{\bibfnamefont{E.}~\bibnamefont{del Valle}},
  \bibnamefont{and} \bibinfo{author}{\bibfnamefont{C.}~\bibnamefont{Tejedor}},
  \bibinfo{journal}{Phys. Rev. Lett.} \textbf{\bibinfo{volume}{101}},
  \bibinfo{pages}{083601} (\bibinfo{year}{2008}).

\bibitem[{\citenamefont{Laucht et~al.}(2009)\citenamefont{Laucht, Hauke,
  Villas-B\^oas, Hofbauer, B\"ohm, Kaniber, and Finley}}]{laucht09b}
\bibinfo{author}{\bibfnamefont{A.}~\bibnamefont{Laucht}},
  \bibinfo{author}{\bibfnamefont{N.}~\bibnamefont{Hauke}},
  \bibinfo{author}{\bibfnamefont{J.~M.} \bibnamefont{Villas-B\^oas}},
  \bibinfo{author}{\bibfnamefont{F.}~\bibnamefont{Hofbauer}},
  \bibinfo{author}{\bibfnamefont{G.}~\bibnamefont{B\"ohm}},
  \bibinfo{author}{\bibfnamefont{M.}~\bibnamefont{Kaniber}}, \bibnamefont{and}
  \bibinfo{author}{\bibfnamefont{J.~J.} \bibnamefont{Finley}},
  \bibinfo{journal}{Phys. Rev. Lett.} \textbf{\bibinfo{volume}{103}},
  \bibinfo{pages}{087405} (\bibinfo{year}{2009}).

\end{thebibliography}

\end{document}